\documentclass[a4paper]{article}
\usepackage[widelayout,hyperrefblack,copyright]{./research17}

\renewcommand{\mycopyright}{Yang, Li, Tang, Wang, \todaylong}

\usepackage{threeparttable}
\usepackage{diagbox}
\usepackage{array}
\usepackage{boldline}
\usepackage{caption}
\usepackage{bbm}
\usepackage{algorithmicx}
\usepackage{algpseudocode}

\allowdisplaybreaks
\usetikzlibrary{datavisualization} 

\usepackage{pgfplots}
\usepgfplotslibrary{units}
\usetikzlibrary{spy,backgrounds}
\usetikzlibrary{decorations.markings}
\usepackage{pgfplotstable}
\usepackage{appendix}

\renewcommand{\indicatorfunction}{\mathop{}\!\mathbbm{1}}

\newcommand{\OSNR}{\mathsf{OSNR}}
\newcommand{\R}{R}

\providecommand{\keywords}[1]{\textbf{Index terms ---} #1}

\theoremstyle{definition}

\providecommand{\keywords}[1]{\textbf{Index terms ---} #1}
\makeatletter
\def\thanks#1{\protected@xdef\@thanks{\@thanks
		\protect\footnotetext{#1}}}
\makeatother

\begin{document}

\title{Tradeoff between Diversity and Multiplexing Gains in Block Fading Optical Wireless Channels}

\author{Sufang Yang, Longguang Li$^\dagger$, Haoyue Tang, Jintao Wang   
\thanks{This work is supported by the National Natural Science Foundation of China under Grant No. 62101192 and Shanghai Sailing Program under Grant No. 21YF1411000. \textit{(Corresponding author: L. Li)}}
\thanks{
Sufang~Yang, Haoyue Tang, and Jintao Wang are with the Department of Electronic Engineering, Tsinghua University, Beijing 100084, China (e-mail: ysf20@tsinghua.org.cn, thy17@mails.tsinghua.edu.cn, wangjintao@tsinghua.edu.cn). }
\thanks{Longguang Li is with Department of Communication and Electronic Engineering, East China Normal University, Shanghai 200241, China (e-mail: lgli@cee.ecnu.edu.cn).}
}
\date{}

\maketitle
\begin{abstract}
The diversity-multiplexing tradeoff (DMT) provides a fundamental performance metric for different multiple-input multiple-output (MIMO) schemes in wireless communications. In this paper, we explore the block fading optical wireless communication (OWC) channels and characterize the DMT in the presence of both optical peak- and average-power constraints. Three different fading distributions are considered, which reflect different channel conditions. In each channel condition, we obtain the optimal DMT when the block length is sufficiently large, and we also derive the lower and upper bounds of the DMT curve when the block length is small. These results are dramatically different from the existing DMT results in radio-frequency (RF) channels. These differences may be due to the fact that the optical input signal is real and bounded, while its RF counterpart is usually complex and unbounded. 

\end{abstract}

\section{Introduction}
\label{sec:introduction}

As an important complement to conventional RF communication, OWC \cite{Vincent2006,Lubin2009} significantly improves the rate performance and offers an ideal solution to the spectrum scarcity in existing wireless communication systems. Recently it has been considered as a promising technique in future 6G \cite{Song2015,Mohamed2016,Hemani2017}. Most current OWC systems adopt the so-called intensity-modulation and direct-detection (IM-DD) transmission scheme because of its simplicity and low-cost deployment. In such a scheme, the transmitter modulates the intensity of optical signals coming from light-emitting diodes (LEDs), and the receiver measures incoming optical intensities by photodetectors \cite{C1997,Sian2009}. As a consequence, transmit signals are proportional to optical intensities, and hence are real and nonnegative, fundamentally different from their RF counterparts. Furthermore, considering {\color{black}safety} reasons and hardware limitations, the peak and average {\color{black}optical} powers of {\color{black}transmitting} signals typically have to be restricted. 

The OWC channels are more sensitive to environment fluctuations than the traditional RF channels due to the IM-DD transmission scheme. Particularly in the medium- and long-range OWC, the fluctuations caused by atmospheric turbulence degrade the quality of the communication links severely. They change the temperature and pressure of the atmosphere, and eventually lead to the refractive index variations of the path links \cite{Zhu2002,Andrews2005,korotkova2021}. To combat the channel fading induced by atmospheric fluctuations in the OWC systems, the widely adopted MIMO technique plays a pivotal role~\cite{AROGYASWAMI2004,LuLu2014,Erik2014,Robert2016}. It provides many independent transmission paths available at the transmitting or receiving ends to effectively alleviate the atmospheric turbulence path loss \cite{Alamouti1998}. Another benefit provided by utilizing MIMO techniques is the significant improvement on spectral efficiency when compared with single-antenna systems \cite{Tarokh1998}.

In this paper, we investigate the OWC-MIMO system from a DMT perspective \cite{Lizhong_Zheng2003,Narasimhan2006,Leizhao2007}. The optimal DMT characterizes the maximal achievable diversity gain at a fixed multiplexing gain, which provides a fundamental metric for comparison between different MIMO transmission schemes, and has triggered extensive research, such as deriving the optimal DMT curves in different channel models \cite{Ali2017,Hatef2016,Daniel2015}, designing various optimal DMT-approaching coding schemes \cite{Usman2018,Tao2020,Christos2015}. 

Most of the existing DMT results focus on the traditional RF systems \cite{Lizhong_Zheng2003,Narasimhan2006,Leizhao2007,Daniel2015,Hatef2016,Ali2017}, among which Zheng and Tse presented the classic optimal tradeoff curve for Rayleigh fading channels \cite{Lizhong_Zheng2003}. They showed that when the channel block length satisfies $l\geq l_\textnormal{th}^\textnormal{ZT}= \nt+\nr-1$, the optimal DMT curve can be exactly characterized as
\begin{IEEEeqnarray}{c}
    d^\star_{\textnormal{ZT}}(r) = (\nt-r)(\nr-r), \label{d(r) by zheng}
\end{IEEEeqnarray}
where $d^\star_{\textnormal{ZT}}(r)$ represents the maximum diversity gain achieved at the given multiplexing gain $r$, and $\nt$ and $\nr$ represent the number of transmit and receive antennas, respectively. Moreover, they proposed the bounds on DMT curve at the small block length regime ($l < l_\textnormal{th}^\textnormal{ZT}$), if $r \leq r_{1}$, with $r_1 = \nr- \lceil \frac{l-|\nt-\nr|-1}{2} \rceil$, the maximum diversity gain is bounded by
\begin{IEEEeqnarray}{rCl}
    d^\star_{\textnormal{ZT}}(r) &\leq& (\nt-r)(\nr-r),\label{du by zheng1}\\
    d^\star_{\textnormal{ZT}}(r) &\geq& -l(r-r_{1}) + (\nt-r_{1})(\nr-r_{1}), \label{du by zheng2}
\end{IEEEeqnarray}
otherwise the optimal DMT curve agrees with \eqref{d(r) by zheng}. This work established the theoretical framework for the DMT characterizations, and many similar results in other channels, such as Rician, Nakagami and log-normal channels are also obtained afterward \cite{Leizhao2007,Shin2008,Safari2008,Hatef2016}. 

Although there has been extensive research on the DMT in RF systems, few studies tap into characterizing the optimal DMT for the OWC systems. Some recent results have shown that many derivations in RF systems are not applicable for OWC systems due to the unique IM-DD transmission scheme \cite{Safari2008,Jaiswal2019,Pranav2020}. The existing work closely related to ours was done by Jaiswal and Bhatnagar~\cite{Jaiswal2019}. They considered the real-valued requirements for the input, and then proposed that if the channel block length satisfies $l\geq l_\textnormal{th}^\textnormal{JB}=\nt+\nr-1$, the optimal DMT curve for negative exponential channel is given by
\begin{IEEEeqnarray}{c}
    d^\star_{\textnormal{JB}}(r) = \frac{1}{2}d^\star_{\textnormal{ZT}}(r). \label{d(r) by jaiswal}
\end{IEEEeqnarray}
Also, under a small block length, they found analogous DMT bounds as in~\eqref{du by zheng1}, \eqref{du by zheng2}.

The one-half factor in~\eqref{d(r) by jaiswal} reflects the loss of half of the degree of freedom compared with RF channels, which is due to the fact that optical inputs need to be real-valued. However, as mentioned before, inputs in OWC channels represent optical intensities, and hence their values must also be nonnegative. In fact, it is the nonnegativity of the optical inputs that significantly complicates the analysis of performance limits in OWC channels. 
Hence, directly applying half of the traditional RF MIMO capacity formula to calculate diversity gain in OWC channels cannot be theoretically justified. Also, the nonnegativity of optical inputs implies that the commonly used two-sided Gaussian random codes in RF channels are no longer admissible in OWC channels. Furthermore, due to the optical intensity inputs, the power constraints imposed on the inputs need to be described differently. All the above issues indicate the existing DMT results in RF channels cannot be directly extended here. Hence, how to characterize the optimal DMT in practical OWC channels still remains an open problem. 

In this paper, we first investigate the optimal DMT by fully considering practical optical input constraints in three different fading channels. We first restrict the optical inputs to be real-valued and nonnegative. Also, a peak- and an average-power constraints are imposed on the inputs. Then we use negative exponential, gamma-gamma, and log-normal distributions to model the channel with atmospheric turbulence fluctuation from strong to weak intensities. For each channel condition, we establish optimal DMT in different block length regimes. It turns out that our derived results are fundamentally different from the above existing results, which may more precisely reflect the fundamental limits of practical OWC systems.  

Specifically, the main contributions in this paper are as follows.
\begin{itemize}
\item \emph{Bounds on Instantaneous Capacity:} By using a truncated exponential random coding argument and further applying the generalized entropy power inequality (GEPI) \cite{Ram_Zamir1993}, we first derive a lower bound on instantaneous capacity. Then, an upper bound is established by some algebraic manipulations on the asymptotic capacity in \cite[Theorem~21]{longguang_li2020}. These bounds are closed-form and proved to be optimal in terms of outage diversity gain.

\item \emph{Exact Characterization of Outage Diversity Gain:} 
With the above new instantaneous capacity bounds, we establish lower and upper bounds on outage diversity gain in general fading channels. Applying these bounds into our considered channels, we precisely characterize the outage diversity gain.

\item \emph{Error Probability Bounds on Truncated Exponential Random Coding:} We propose a truncated exponential random coding scheme, which helps to derive a new upper bound on average error probability. Based on this bound, we derive a tight lower bound on the optimal diversity gain in different fading channels.  

\item \emph{Characterization of Optimal DMT:} 
We characterize the optimal DMT curves in different block length regimes for considered channels. Specifically, if the block length $l \geq \nt-\nr +1$, the optimal diversity gain $d^\star(r)$ can be characterized as
\begin{IEEEeqnarray}{c}
    d^\star(r) = (\nt-\nr + 1)(\nr-r).
\end{IEEEeqnarray}

Otherwise under a small block length ($l < \nt-\nr+1$), the optimal diversity gain can be upper- and lower-bounded by:
\begin{IEEEeqnarray}{rCl}
    d^\star(r) &\leq& (\nt-\nr+1)(\nr-r),\\
    d^\star(r) &\geq& l(\nr-r). 
\end{IEEEeqnarray}
See Theorems~\ref{theorem: negative exponential DMT}, \ref{theorem: gamma-gamma DMT}, and \ref{theorem: log-normal DMT} in detail. 
\end{itemize}

The paper is organized as follows. We end the introduction with a few notational conventions. Section~\ref{sec:channel-model} describes in detail the investigated channel model. In Section~\ref{sec-outage probability bounds}, we present new upper and lower bounds on outage diversity gain. Section~\ref{sec-Random Coding Error Analysis} characterizes the average error probability with the truncated exponential random coding. The optimal DMTs of negative exponential, gamma-gamma, and log-normal channels are characterized in Sections~\ref{sec:Negative Exponential Atmospheric Turbulence}, \ref{sec:Gamma-Gamma Atmospheric Turbulence} and \ref{sec:Log-Normal Atmospheric Turbulence}. Numerical examples are included in Section~\ref{sec: Numerical Examples}. Most of the proofs are in the appendices.

\textbf{Notation:} Random variables and matrices are boldfaced, e.g., $\rv{h}$ and $\mat{X}$, while their realizations are typeset in $h$ and $\mathrm{X}$, respectively. $\mat{X}_j$ denotes the $j$th column of matrix $\mat{X}$, and $\rv{x}_{ij}$ or $[\mat{X}]_{ij}$ denotes the $i$th row and $j$th entry of $\mat{X}$.  Sets are typeset in a special font, e.g., $\mathcal{A}$. Differential Entropy is denoted by $\hh(\cdot)$, and mutual information by $\II(\cdot ; \cdot)$. $\|\cdot\|_{1}$ and $\|\cdot\|_\mathsf{F}$ denote the $\mathcal{L}_{1}$- and Frobenius-norm, respectively. $\log (\cdot)$ denotes the logarithm to the base of $e$. The expectation of a random variable is denoted by $ \E{\cdot}$, and variance by $\Var{\cdot}$. $\mathcal{R}^{m \times n}$ ($\mathcal{R}_+^{m \times n}$) denotes real (nonnegative) valued set. We denote $x^+ \eqdef \max \{0,x\}$, and use symbol $\doteq$ to denote \emph{exponential equality}, i.e., $g(x)\doteq x^b$ indicates $\lim _{x \rightarrow \infty} \frac{\log g(x)}{\log x}=b$, and $\dot{\geqslant} , \dot{\leqslant}$ are similarly defined.

\section{Channel Model}
\label{sec:channel-model}
Consider a MIMO channel with $\nt$ LEDs and $\nr$ photodetectors. The channel output is given by\footnote{For simplicity, we assume the photoelectric coefficient for the photodetector is $1$.}
\begin{IEEEeqnarray}{c}
    \mathbf{Y}=\mathbf{H}\mathbf{X}+\mathbf{Z}, \label{channel model}
\end{IEEEeqnarray}
where $\mathbf{X} \in \mathcal{R}_+^{\nt \times l}$ denotes the channel input, with $l$ being the block length; where $\mathbf{Z}\in \mathcal{R}^{\nr \times l}$ denotes the channel noise, whose entries are independent and identically distributed (i.i.d.) inside and across blocks; and where $\mathbf{H} \in \mathcal{R} _+^{\nr \times \nt}$ denotes the channel matrix, and its entries remain constant inside one block, and i.i.d. across blocks.

The entry $\rv{h}_{ij}$ in $\mathbf{H}$ represents the nonnegative and real-valued gain from $j$th transmit antenna to $i$th receive antenna, and it depends on two factors: deterministic distance attenuation and random atmospheric turbulence loss. Hence, $\rv{h}_{ij}$ can be formulated as
\begin{IEEEeqnarray}{c}
    \rv{h}_{i j}=h_{i j}^{d} \rv{h}_{i j}^{r}, \label{h_ij component}
\end{IEEEeqnarray}
where $h_{i j}^{d}=e^{-\nu d_{i j}}$ represents the deterministic part with parameter $\nu$ characterizing the transmission environment and $d_{i j}$ being the transmission distance; and where $\rv{h}_{i j}^{r}$ represents the random part characterizing the atmospheric turbulence intensity. In this paper, we consider three different distributions of the random atmospheric turbulence $\rv{h}_{i j}^{r}$, which cover the turbulence fluctuation regimes from strong to weak intensities, and they are

\begin{itemize}
\item \emph{Negative exponential distribution:}
    \begin{IEEEeqnarray}{rCl}
    {f}({h}_{i j}^{r}) = e^{-{h}_{i j}^{r}};\label{negative exponential pdf}
    \end{IEEEeqnarray}
    
\item \emph{Gamma-gamma distribution:}
    \begin{IEEEeqnarray}{rCl}
        {f}({h}_{i j}^{r})=\frac{2(\rho_{1}\rho_{2})^{\frac{\rho_{1}+\rho_{2}}{2}}}{\Gamma(\rho_{1})\Gamma(\rho_{2})}({h}_{i j}^{r})^{\frac{\rho_{1}+\rho_{2}}{2}-1}K_{\rho_{1}-\rho_{2}}(2\sqrt{\rho_{1}\rho_{2} {h}_{i j}^{r}}), \label{gamma gamma pdf}
    \end{IEEEeqnarray}
where $\rho_{1}$ and $\rho_{2}$ denote the irradiance fluctuation parameters with $\rho_{2} <\rho_{1}$; where $\Gamma(\cdot)$ denotes the Gamma function with $\Gamma(z)=\int_{0}^{\infty} x^{z-1} e^{-x} dx$; and where $K_{\tau}(\cdot)$ denotes the modified Bessel function of the second kind with ${\tau}$ being the order \cite{weisstein2002,Bhatnagar2015};

\item \emph{Log-normal distribution:}
    \begin{IEEEeqnarray}{rCl}
        {f}({h}_{i j}^{r})=\frac{1}{{h}_{i j}^{r}\sqrt{2\pi{\sigma}_l^2}}\exp\left(-\frac{(\log({h}_{i j}^{r})-\mu_l)^2}{2{\sigma}_l^2}\right), \label{log normal pdf}
    \end{IEEEeqnarray}
where $\mu_l$ and ${\sigma}_l$ denote the expectation and variance of $\log(\rv{h}_{i j}^{r})$, respectively.
\end{itemize}

Considering the limited dynamic working range of the light emitters and practical illumination requirements for the modulated optical sources, both peak- and average-power constraints are imposed on the channel input, i.e.,  
\begin{IEEEeqnarray}{l}
    \subnumberinglabel{eq:constraints}
    \mathrm{P}\Bigl[\rv{x}_{i j}> \amp \Bigr] =  0,  \ \forall i \in \{1,\ldots,\nt\}, \ \forall j \in \{1,\ldots,l\}, \quad \label{peak-power constraint}\\
    \frac{1}{l} \BigE{\sum_{j=1}^{l}\parallel\vect{X}_j\parallel_1} \leq\EE,  
    \label{average-power constraint}
\end{IEEEeqnarray}
where $\vect{X}_j$ denotes the $j$th column input vector; where $\amp$ represents the allowed maximum optical power by each antenna; and where $\EE$ denotes the total average optical power allowed across all antennas. The ratio between the allowed average power and the allowed peak power is denoted by
\begin{IEEEeqnarray}{c}
    \alpha \triangleq \frac{\EE}{\amp},\label{ratio eta}
\end{IEEEeqnarray}
where $\alpha \in (0,\frac{\nt}{2}]$, and is fixed in the paper.

Since information is carried on the intensity of the optical signal, we adopt the definition of optical signal-to-noise ratio (OSNR) \cite{Riedl2001,Faruk2014,Kamran2020} as follows:
\begin{IEEEeqnarray}{c}
    \OSNR=\frac{\EE}{{\sigma_n}}.
\end{IEEEeqnarray}

We further present some useful concepts and definitions in terms of OSNR. More details can be seen in \cite{Lizhong_Zheng2003}.

Given a transmission scheme, an outage occurs when the mutual information of this channel can not support the target rate $\R$, and we denote \begin{IEEEeqnarray}{rCl}
    \set{C}
    &\triangleq& \left\{\matt{H}:\II(\mat{X}_j;\mat{Y}_j|\mat{H}=\matt{H})<\R\right\}\label{set C definition}
\end{IEEEeqnarray}
 as the outage event in terms of $\mat{H}$.

Among all possible transmission schemes, the outage probability at time index $j$ is defined as \cite{Lizhong_Zheng2003}
\begin{IEEEeqnarray}{C}
    \mathrm{P}_{\textnormal{out}}(\OSNR) \eqdef \min_{{f}(\matt{X}_j) \textnormal{ satisfying } \eqref{eq:constraints}} \mathrm{P}\bigl[ \II(\vect{X}_j;\vect{Y}_j|\mat{H}) \leq \R \bigr], \label{Pout definition}
\end{IEEEeqnarray}
where $\vect{X}_j$ and $\vect{Y}_j$ denote the transmit and receive vectors at time $j$, respectively; and where ${f}(\matt{X}_j)$ denotes the input distribution of $\mat{X}_j$. In the existing literature~\cite{Ahmed2007,Hatef2016}, the mutual information term in~\eqref{Pout definition} is also called instantaneous capacity. 

Now we briefly define the following diversity and multiplexing gains in terms of OSNR that will be used in the rest of the paper. 
\begin{definition}
A transmission scheme is said to achieve multiplexing gain $r$, outage diversity gain $d_{\textnormal{out}}(r)$, and diversity gain $d(r)$ if the rate $R(\OSNR)$ satisfies
\begin{IEEEeqnarray}{l}
      \lim _{\OSNR \rightarrow \infty} \frac{\R(\OSNR)}{\log (\OSNR)} = r,
\end{IEEEeqnarray}
the outage probability $\mathrm{P}_{\textnormal{out}}(\OSNR)$ satisfies 
\begin{IEEEeqnarray}{C}
    -\lim_{\OSNR\rightarrow\infty}{\frac{\log{\mathrm{P}_{\textnormal{out}}(\OSNR)}}{\log(\OSNR)}} = d_{\textnormal{out}}(r), \label{d_out definition}
\end{IEEEeqnarray}    
and average error probability $\mathrm{P}_{\textnormal{e}}(\OSNR)$ satisfies
\begin{IEEEeqnarray}{l}
 -\lim_{\OSNR\rightarrow\infty}{\frac{\log{\mathrm{P}_{\textnormal{e}}(\OSNR)}}{\log(\OSNR)}} = d(r).\label{d definition}
    \end{IEEEeqnarray}

For each $r$, we define $d^\star(r)$ as the supremum of the diversity gain achieved over all schemes at the data rate $\R(\OSNR)$. 
\end{definition}    

\section{Outage Probability Analysis}
\label{sec-outage probability bounds}
This section presents the new outage probability bounds, which are crucial in the following derivations of the optimal DMTs.  

To estimate the outage probability in~\eqref{Pout definition}, we first need to characterize the instantaneous capacity. In the presence of the peak- and average-power constraints in \eqref{peak-power constraint} and \eqref{average-power constraint}, there are no existing results in current literature applicable here. We present new lower and upper bounds, which are closed-form and sufficiently tight at high $\OSNR$. 

The lower bound is derived by using a truncated exponential random coding argument. For any time index $j \in \{1,2,\ldots,l \}$, denote $\vect{X}_j = \trans{[\mat{X}_{j1},\mat{X}_{j2},\ldots,\mat{X}_{ j\nt}]}$, and let the entries of $\vect{X}_j$ be i.i.d. according to the following truncated exponential distribution:
\begin{IEEEeqnarray}{rCl}
   {f}(\matt{X}_{ji})=\frac{1}{{\amp}} \cdot \frac{\mu}{1-e^{-\mu}} \cdot e^{-\frac{\mu\matt{X}_{ji}}{\amp}},\quad \forall i \in \{1,2,\ldots, \nt \}, \label{exponential distribution2}
\end{IEEEeqnarray}
where $\mu$ is a parameter satisfying
\begin{IEEEeqnarray}{rCl}
    \frac{1}{\mu}-\frac{e^{-\mu}}{1-e^{-\mu}}=\frac{\alpha}{\nt}. \label{eq:mu2}
\end{IEEEeqnarray}

The achievable rate by this truncated exponential distribution can serve as a natural lower bound on the instantaneous capacity. Applying the GEPI in~\cite{Ram_Zamir1993} we derive the following lower bound, whose rigorous proof is shown in~Appendix~\ref{app: proof of mutual information bounds}.
\begin{proposition}[Lower Bound]
\label{proposition: mutual information bounds}
Given a MIMO-OWC channel in \eqref{channel model}, we have
    \begin{IEEEeqnarray}{rCl}
        \II(\vect{X}_j;\vect{Y}_j|\mat{H}) &\geq& \frac{1}{2} \log \left(1+L_{l} (\OSNR)^{2\nr}\left|\mat{H}\trans{\mat{H}}\right|\right), \label{lower bound on mutual information}
    \end{IEEEeqnarray}
where 
\begin{equation}
L_{l}=\left(\frac{2\sigma_n^2}{\pi\alpha^2 e}\right)^{\nr} \left( \frac{1-e^{-\mu}}{\mu} 2^{-\frac{\mu e^{-\mu}}{1-e^{-\mu}}}\right)^{2\nr}
\end{equation}
with $\mu$ satisfying~\eqref{eq:mu2}.
\end{proposition}

The following upper bound is derived by first assuming the channel state information available at the transmitter, and then by some algebraic manipulations on the existing asymptotic capacity in~\cite[Theorem~21]{longguang_li2020}. 
\begin{proposition}[Upper Bound]
\label{proposition: mutual information bounds2}
Given a MIMO-OWC channel in \eqref{channel model}, we have
\begin{IEEEeqnarray}{rCl}
        \II(\vect{X}_j;\vect{Y}_j|\mat{H}) &\leq& \frac{1}{2} \log \left(L_{u} (\OSNR)^{2\nr}\left|\mat{H}\trans{\mat{H}}\right|\right),\label{upper bound on mutual information}
    \end{IEEEeqnarray}
where 
\begin{IEEEeqnarray}{rCl}
    L_{u}={{\nt}\choose{\nr}}\left(\frac{\sigma_n^2}{2\pi\alpha^2e}\right)^{\nr}.
\end{IEEEeqnarray}

\end{proposition}

\begin{remark}
Note that $L_{u}$ and $L_{l}$ in Proposition~\ref{proposition: mutual information bounds} and~\ref{proposition: mutual information bounds2} are constants independent of parameters $\OSNR$ and $\mat{H}$, which play no role in the following derivations related to the OSNR exponent. 
\end{remark}

For a given channel realization $\matt{H}$, we denote ${\lambda}_i, \,\, \forall  i=\{1,\ldots,\nr\}$, as eigenvalues of $\matt{H}\trans{\matt{H}}$ with $\lambda_i$ arranged in an increasing order. It is obvious they are real and nonnegative. We further express ${\lambda}_i=(\OSNR)^{-{a}_i}$ for $i\in\{1,\cdots,\nr\}$. For the notational convenience, we denote vector $a\eqdef[a_1,\cdots,a_{\nr}]$ and vector ${\lambda} \eqdef[{\lambda}_1,\cdots,{\lambda}_{\nr}]$.\footnote{It should be noted that the random forms corresponding to $a$ and ${\lambda}$ are denoted as $\boldsymbol{a}$ and $\bm{\lambda}$.}

Decompose $\matt{H}$ as $\matt{UDQ}$, where $\matt{U}\in\set{R}^{\nr\times\nr}$ and $\matt{Q}\in\set{R}^{\nr\times\nt}$ are orthogonal matrices, and $\matt{D}$ is a diagonal matrix with $\matt{D}=\operatorname{diag}[(\OSNR)^{-\frac{a_{1}}{2}}, \ldots, (\OSNR)^{-\frac{a_{\nr}}{2}}]$. More details about the matrix decomposition can be found in Appendix~\ref{proof of f(a)}.

For a given distribution of $\mat{H}$, by \cite{Alan1988} we prove in~Appendix~\ref{proof of f(a)} that the distribution of vector $\boldsymbol{a}$ can be estimated as
\begin{IEEEeqnarray}{c}
    {f}(a) \doteq (\OSNR)^{-\sum_{i=1}^{\nr} \frac{\nt-\nr+2i-1}{2}a_i} \times \iint {f}(\matt{UDQ}) d \matt{Q} d \matt{U}, \label{f(a) definition}
\end{IEEEeqnarray}
with ${f}(\matt{UDQ})$ being the distribution of $\mat{H}$. 

Now we are ready to present the main theorem with proof in Appendix~\ref{app-proof of new bounds on outage probability}.

\begin{theorem}[Outage Probability Bounds]
\label{proposition-outage probability bounds}
Given a channel in~\eqref{channel model}, we have
\begin{IEEEeqnarray}{rCl}
  \int_{\set{B}} {f}(a) d a  \,\,  \dot{\leqslant} \,\, \mathrm{P}_{\textnormal{out}}(\OSNR) \,\,
   &\dot{\leqslant}&\, \int_{\set{A}} {f}(a) d a, \label{integration with A}
\end{IEEEeqnarray}
where set $\set{A}$ is defined as
\begin{IEEEeqnarray}{rCl}
    \set{A} &=& \left\{a:(2 \nr-\sum_{i=1}^{\nr} a_{i})^{+} \leq 2 r\right\},\label{set A definition}
\end{IEEEeqnarray}
and set $\set{B}$ as
\begin{IEEEeqnarray}{rCl}
    \set{B} &=& \left\{a:2 \nr-\sum_{i=1}^{\nr} a_{i} \leq 2 r\right\};\label{set B definition}
\end{IEEEeqnarray}
where $f(a)$ is defined as in~\eqref{f(a) definition}.
\end{theorem}
\medskip

\section{Random Coding Error Analysis}
\label{sec-Random Coding Error Analysis}
This section presents the diversity order of truncated exponential random coding, which serves as a lower bound on the optimal diversity order. 

Consider a random code with codewords i.i.d.$\ $as in~\eqref{exponential distribution2}, and the code rate is $\R=r\log(\OSNR)$. The decoder applies the maximum-likelihood (ML) method to detect the sent message. Now consider a channel realization $\mat{H} = \matt{H}$, then the conditional error probability can be upper bounded by the following inequality, whose proof is postponed to Appendix~\ref{sec-Proof of Proposition random coding analysis}.
\begin{IEEEeqnarray}{rCl}
\mathrm{P}(\textnormal{error}|\mat{H}=\matt{H})
    &\leq& (\OSNR)^{lr} \prod_{i=1}^{\nr}\left(1+\frac{g(\alpha,\nt)}{2}(\OSNR)^2\lambda_i\right)^{-\frac{l}{2}},\label{condition error probability2}
\end{IEEEeqnarray}
where 
\begin{equation}
g(\alpha,\nt)= {\frac{1}{\mu^2}-\frac{e^{-\mu}}{(1-e^{-\mu})^2}} 
\end{equation}
with $\mu$ satisfying~\eqref{eq:mu2}.

Now we present the main theorem on the error probability $\mathrm{P}_{\textnormal{e}}(\OSNR)$ of this random coding scheme.
\begin{theorem}\label{theorem: random coding error}
Given a channel in \eqref{channel model}, we have
\begin{IEEEeqnarray}{rCl}
    \mathrm{P}_{\textnormal{e}}(\OSNR)  
    \,\,&\dot{\leqslant}&\,\, (\OSNR)^{-d_\textnormal{out}(r)} + (\OSNR)^{-d_\textnormal{te}(r)}, \label{Pe upper bound}
\end{IEEEeqnarray}
where 
\begin{align}
    (\OSNR)^{-d_\textnormal{te}(r)}\doteq\int_{\mathcal{C}^{c}} f(a) (\OSNR)^{-\frac{l}{2}(\sum_{i=1}^{\nr}(2-a_i)^{+}-2r)} da, \label{d_G definition}
\end{align}
with $\mathcal{C}^c$ denoting the complement of outage set $\mathcal{C}$.
\end{theorem}
\begin{IEEEproof}
We consider the error event conditioned on the channel outage event. Specifically, we bound the error probability by 
\begin{align}
    \mathrm{P}_{\textnormal{e}}(\OSNR) 
    &= \mathrm{P}_{\textnormal{out}}(\OSNR) \mathrm{P}_{\textnormal{e|out}}(\OSNR)+ \mathrm{P}_{\textnormal{e,nout}}(\OSNR) \\
    &\leq \mathrm{P}_{\textnormal{out}}(\OSNR)+\mathrm{P}_{\textnormal{e,nout}}(\OSNR). \label{error probability}
\end{align}
where $\mathrm{P}_{\textnormal{e|out}}(\OSNR)$ denotes the error probability conditioned on outage event; where $\mathrm{P}_{\textnormal{e,nout}}(\OSNR)$ denotes the joint probability of error and no outage events.

By the definition of outage diversity, the first term in the RHS of ~\eqref{error probability} is 
\begin{equation}
   \mathrm{P}_{\textnormal{out}}(\OSNR) \doteq  (\OSNR)^{-d_\textnormal{out}(r)}. 
\end{equation}
Substitute $\lambda_i = (\OSNR)^{-a_i}$ into~\eqref{condition error probability2}, we have 
\begin{IEEEeqnarray}{rCl}
\mathrm{P}(\textnormal{error}|a) \ \dot{\leqslant}\  (\OSNR)^{-\frac{l}{2}(\sum_{i=1}^{\nr}(2-a_i)^{+}-2r)}.\label{}
\end{IEEEeqnarray}
Multiplying with the distribution of $a$, and integrating over set $\set{C}^c$, the second term in the RHS of~\eqref{error probability} is
 \begin{equation}
   \mathrm{P}_{\textnormal{e,nout}}(\OSNR) \doteq  (\OSNR)^{-d_\textnormal{te}(r)}. 
\end{equation}
 The proof is concluded. 
\end{IEEEproof}

\section{DMT Analysis on Negative Exponential Channel}
\label{sec:Negative Exponential Atmospheric Turbulence}
In this section we consider the OWC channel with atmospheric turbulence according to negative exponential distribution. This distribution is commonly used to model the channel with relatively high atmospheric turbulence intensity. We first present results on the outage diversity order, and based on this, we further characterize the optimal DMT.

\subsection{Outage Diversity Gain Characterization} 
\label{subsec-outage diversity gain characterization}
To distinguish parameters from different fading channels, here we denote $l_\textnormal{th}^\textnormal{NE}$, $d_\textnormal{out}^\textnormal{NE}(r)$, $d_\textnormal{te}^\textnormal{NE}(r)$ and $d_\textnormal{NE}^\star(r)$ as the parameters for negative exponential channel. Similar notations will also be used in the following gamma-gamma and log-normal channels. 

As shown in Theorem~\ref{proposition-outage probability bounds}, to estimate the outage diversity order, we first need to calculate the distribution of vector $\rv{a}$ in~\eqref{f(a) definition}. To do this, we first analyze the distribution of $\mat{H}$. Since the entries in $\mat{H}$ are assumed i.i.d., and by \eqref{negative exponential pdf} we expand the distribution of $\mat{H}$ as
\begin{IEEEeqnarray}{rCl}
    f(\matt{H})
    &=& \prod_{i=1}^{\nr} \prod_{j=1}^{\nt}{f}\left({h}_{i j}\right) \label{MIMO p(H)} \\
        &=& \prod_{i=1}^{\nr}\prod_{j=1}^{\nt} \exp \left(-e^{\nu d_{i j}}   {h}_{i j}+\nu d_{i j}\right). \label{whole pdf negative exponential}
\end{IEEEeqnarray}
Now we decompose $\matt{H} =\matt{UDQ}$, and then for any entry $h_{ij} = \left[ \matt{H} \right]_{i j}$, we have
\begin{equation}
h_{ij}=\sum_{k=1}^{\nr}u_{ik} \times (\OSNR)^{-\frac{a_k}{2}} \times q_{kj}, \label{eq:hijexpand}
\end{equation}
 where $u_{i k}=\left[ \matt{U} \right]_{i k}$ and $q_{i k}=\left[ \matt{Q} \right]_{i k}$. Since $a_i$s are arranged in a decreasing order, at high OSNR, the last term $u_{i \nr} \times (\OSNR)^{-\frac{a_{\nr}}{2}} \times q_{\nr j}$ dominates other terms. By the fact that the entries of $\matt{H}$ are nonnegative, we have $u_{i \nr} q_{\nr j}\geq 0$. Hence ${f}\left(\matt{H}\right)$ can be further simplified as
\begin{IEEEeqnarray}{rCl}
    f(\matt{H}) 
    &\doteq& \prod_{i=1}^{\nr}\prod_{j=1}^{\nt} \exp \left(- (\OSNR)^{-\frac{a_{\nr}}{2}}\right). \label{NE $pH(a)$}
\end{IEEEeqnarray}
Substituting \eqref{NE $pH(a)$} into \eqref{f(a) definition}, we get
\begin{IEEEeqnarray}{c}
    f(a) \doteq (\OSNR)^{-\sum_{i=1}^{\nr} \frac{\nt-\nr+2i-1}{2}a_i} \times\prod_{i=1}^{\nr}\prod_{j=1}^{\nt} \exp \left(- (\OSNR)^{-\frac{a_{\nr}}{2}}\right). \label{f(a) definition2}
\end{IEEEeqnarray}

Now we apply Theorem~\ref{proposition-outage probability bounds} to bound the outage probability. Note that when $a_{\nr}<0$, in~\eqref{f(a) definition2} the exponential term $\exp{\left(-(\OSNR)^{-\frac{a_{\nr}}{2}}\right)} \rightarrow 0$, and hence $f(a) \rightarrow 0$. With this observation, we can further reduce the integral domain $\set{A}$ in \eqref{integration with A} to be $\set{A}^{\prime}=\set{A} \cap\left\{a: a_{\nr} \geq 0\right\}$ and $\set{B}$ in \eqref{integration with A} to be $\set{B}^{\prime}=\set{B} \cap\left\{a: a_{\nr} \geq 0\right\}$. Furthermore, when $a_{\nr} \geq 0$, $\exp{\left(-(\OSNR)^{-\frac{a_{\nr}}{2}}\right)} \rightarrow 1 \textnormal{ or } e^{-1}$, and then we can further simplify $f(a)$ in~\eqref{f(a) definition2} as
\begin{IEEEeqnarray}{c}
    f(a) \doteq (\OSNR)^{-\sum_{i=1}^{\nr} \frac{\nt-\nr+2i-1}{2}a_i}. \label{new f(a) negative exponential3}
\end{IEEEeqnarray}
Now by definition~\eqref{Pout definition}, substituting~\eqref{new f(a) negative exponential3} into~\eqref{integration with A}, the outage diversity gain $d_\textnormal{out}^\textnormal{NE}(r)$ at multiplexing order $r$ is bounded by
\begin{IEEEeqnarray}{rCl}
    \inf _{a \in \set{A}^{\prime} } d_\textnormal{out}^\textnormal{NE}(r,a)  \leq d_\textnormal{out}^\textnormal{NE}(r) \leq \inf _{a \in \set{B}^{\prime} } d_\textnormal{out}^\textnormal{NE}(r,a)
     \label{dout bound for negative exponential},
\end{IEEEeqnarray}
where function $d_\textnormal{out}^\textnormal{NE}(r,a)$ is defined as
\begin{IEEEeqnarray}{rCl}
    d_\textnormal{out}^\textnormal{NE}(r,a)=\sum_{i=1}^{\nr} \frac{\nt-\nr+2i-1}{2} a_{i}; \label{dout_ne(r,a)}
\end{IEEEeqnarray}
and where $\set{A}^{\prime}$ is given by
\begin{IEEEeqnarray}{l}
    \set{A}^{\prime}= \Biggl\{a : a_{1}\geq \cdots \geq a_{\nr}\geq 0, \textnormal{ and } (2 \nr-\sum_{i=1}^{\nr} a_{i})^{+} \leq 2r \Biggr\},\label{set A prime}
\end{IEEEeqnarray}
and $\set{B}^{\prime}$ by
\begin{IEEEeqnarray}{l}
    \set{B}^{\prime}= \Biggl\{a : a_{1}\geq \cdots \geq a_{\nr}\geq 0, \textnormal{ and } 2 \nr-\sum_{i=1}^{\nr} a_{i} \leq 2r \Biggr\}.\label{set B prime}
\end{IEEEeqnarray}
\subsection{Optimal DMT Characterization}
\label{subsec-optimal DMT characterization}
We now characterize the optimal DMT of negative exponential channel. In fact, our derived outage probability can serve as a lower bound on average error probability. By Fano's inequality \cite{cover2006}, we can show
\begin{IEEEeqnarray}{c}
    \mathrm{P}_{\textnormal{e}}(\OSNR) \,\, \dot{\geqslant} \,\, (\OSNR)^{-d_\textnormal{out}(r)}.\label{Pe lower bound}
\end{IEEEeqnarray}
Furthermore, note that the error probability bound on truncated exponential coding established in Section~\ref{sec-Random Coding Error Analysis} can serve as an upper bound on average error probability. Hence
\begin{IEEEeqnarray}{rCl}
    \mathrm{P}_{\textnormal{e}}(\OSNR)  
  \,\,  &\dot{\leqslant}& \,\, (\OSNR)^{-d_\textnormal{out}(r)} + (\OSNR)^{-d_\textnormal{te}(r)}. \label{Pe upper bound 2}
\end{IEEEeqnarray}

Comparing the bounds in~\eqref{Pe lower bound} and \eqref{Pe upper bound 2}, we have the following observations:
\begin{itemize}
    \item If $d_\textnormal{te}(r) \geq d_\textnormal{out}(r)$, $d_\textnormal{out}(r)$ becomes the dominant term in the RHS of~\eqref{Pe upper bound 2}. Combined with \eqref{Pe lower bound}, the optimal diversity gain $d^{\star}(r)$ is equal to the outage diversity gain, i.e., 
    \begin{IEEEeqnarray}{c}
       d^{\star}(r) = d_\textnormal{out}(r).\label{e6-1}
    \end{IEEEeqnarray}
    \item If $d_\textnormal{te}(r) < d_\textnormal{out}(r)$, $d_\textnormal{te}(r)$ turns to be the dominant term in the RHS of~\eqref{Pe upper bound 2}. In this case, we can simply bound $d^{\star}(r)$ as
    \begin{IEEEeqnarray}{rCl}
       d_\textnormal{te}(r) \leq d^\star(r) \leq d_\textnormal{out}(r).\label{e6-2}
    \end{IEEEeqnarray}
\end{itemize}
With the above two observations, we only need to calculate $d_\textnormal{out}(r)$ and $d_\textnormal{te}(r)$. The optimal DMT of negative exponential channel is characterized in the following theorem. 

\begin{theorem}[DMT of Negative Exponential Channel]
\label{theorem: negative exponential DMT}
Given a channel with distribution in \eqref{whole pdf negative exponential}, if $l\geq l_\textnormal{th}^\textnormal{NE} = \nt-\nr+1$, the optimal diversity order is given by
\begin{IEEEeqnarray}{rCl}
    d_\textnormal{NE}^\star(r) = (\nt-\nr+1)(\nr-r),\label{eq52}
\end{IEEEeqnarray}
otherwise, 
\begin{IEEEeqnarray}{rCl}
 l(\nr-r)  \leq   d_\textnormal{NE}^\star(r) \leq (\nt-\nr+1)(\nr-r). \label{eq53}
\end{IEEEeqnarray}
\end{theorem}
\begin{IEEEproof}
We first calculate $d_\textnormal{out}^\textnormal{NE}(r)$ and $d_\textnormal{te}^\textnormal{NE}(r)$, and then characterize the optimal diversity order $d_\textnormal{NE}^\star(r)$.

By linear optimization, it is straightforward to get that the infimums of $d_\textnormal{out}^\textnormal{NE}(r,a)$ over feasible regimes $\set{A}^{\prime}$ and $\set{B}^{\prime}$ are both achieved at $a_\textnormal{out}^\star=[2(\nr-r),0,\ldots,0]$ in \eqref{dout bound for negative exponential}, and the infimums match. Hence the outage diversity gain is given by
\begin{IEEEeqnarray}{c}
    d_\textnormal{out}^\textnormal{NE}(r)=(\nt - \nr+ 1)(\nr - r).\label{dout negative exponential}
\end{IEEEeqnarray}

 Now we calculate $d_\textnormal{te}^\textnormal{NE}(r)$. By reducing the integral domain $\set{C}^c$ in \eqref{d_G definition} to $\set{A}^c$, and enlarging the integral domain $\set{C}^c$ in \eqref{d_G definition} to $\set{B}^c$, we bound $d_\textnormal{te}^\textnormal{NE}(r) $ as
\begin{IEEEeqnarray}{rCl}
    \inf_{\left(\set{B}^c\right)^{\prime}}
    d_\textnormal{te}^\textnormal{NE}(r,a) \leq d_\textnormal{te}^\textnormal{NE}(r) 
    &\leq&\inf_{\left(\set{A}^c\right)^{\prime}}
    d_\textnormal{te}^\textnormal{NE}(r,a) , \label{new expression of dG for ne channel}
\end{IEEEeqnarray}
where function $d_\textnormal{te}^\textnormal{NE}(r,a)$ is given by
\begin{IEEEeqnarray}{rCl}
     d_\textnormal{te}^\textnormal{NE}(r,a)=\sum_{i=1}^{\nr} \frac{\nt-\nr+2i-1}{2}a_i +
     \frac{l}{2}\biggl(\sum_{i=1}^{\nr}\left(2-a_i\right)^+-2r\biggr); \label{dte(r,a) for ne channel}
\end{IEEEeqnarray}
and where $(\set{A}^c)^{\prime}$ is given by
\begin{IEEEeqnarray}{L}
    (\set{A}^c)^{\prime}= \Biggl\{a : a_{1}\geq \cdots \geq a_{\nr}\geq 0, \textnormal{ and } (2 \nr-\sum_{i=1}^{\nr} a_{i})^{+} \geq 2r \Biggr\},\label{Ac prime}
\end{IEEEeqnarray}
and $(\set{B}^c)^{\prime}$ by
\begin{IEEEeqnarray}{L}
    (\set{B}^c)^{\prime}= \Biggl\{a : a_{1}\geq \cdots \geq a_{\nr}\geq 0, \textnormal{ and } 2 \nr-\sum_{i=1}^{\nr} a_{i} \geq 2r \Biggr\}.\label{Bc prime}
\end{IEEEeqnarray}

Compared with \eqref{dout_ne(r,a)}, we can see $d_\textnormal{te}^\textnormal{NE}(r,a)$ in \eqref{dte(r,a) for ne channel} is also linear with respect to vector $a$, but with an extra term $\frac{l}{2}\bigl(\sum_{i=1}^{\nr}\left(2-a_i\right)^+-2r\bigr)$. Given a block length $l$, we can still analyze the infimum of $d_\textnormal{te}^\textnormal{NE}(r,a)$ over $(\set{A}^c)^{\prime}$ or $(\set{B}^c)^{\prime}$ through linear optimization. Here, we show that the infimums in \eqref{new expression of dG for ne channel} also match the same optimal point $a_\textnormal{te}^\star$. In the following, we characterize $a_\textnormal{te}^\star$ under different block length $l$:

\begin{itemize}
    \item when $l < \nt - \nr + 1$, the optimal $a_\textnormal{te}^\star$ is given by
    \begin{IEEEeqnarray}{rCl}
    \label{eq:lleqlth}
        a_i^\star =0, \textnormal{ for } i=1,\cdots,\nr;\label{eq1: optimal ate}
    \end{IEEEeqnarray}    
    
    \item when $\nt - \nr + 2k -1 \leq l < \nt - \nr + 2(k+1)-1  $, $\forall k \in \{ 1,\cdots, \nr-r\}$, the optimal $a_\textnormal{te}^\star$ is given by
    \begin{IEEEeqnarray}{rCl}
    \subnumberinglabel{eq2: optimal ate}
     a_i^\star & = & 2, \textnormal{ for } i=1,\cdots,k, \\
     a_i^\star & = & 0, \textnormal{ for } i=k+1,\cdots,\nr; 
    \end{IEEEeqnarray}
    
    \item when $ l \geq \nt - \nr + 2((\nr-r)+1) -1 $, the optimal $a_\textnormal{te}^\star$ is given by
    \begin{IEEEeqnarray}{rCl}
    \subnumberinglabel{eq3: optimal ate}
     a_i^\star & = & 2, \textnormal{ for } i=1,\cdots,\nr-r, \\
     a_i^\star & = & 0, \textnormal{ for } i=\nr-r+1,\cdots,\nr.
    \end{IEEEeqnarray}
\end{itemize}
Substituting $a_\textnormal{te}^\star$ into \eqref{new expression of dG for ne channel}, we obtain that if $l<\nt-\nr+1$, then
\begin{IEEEeqnarray}{rCl}
    d_\textnormal{te}^\textnormal{NE}(r) 
    &=& l(\nr-r). \label{eq:dte}
\end{IEEEeqnarray}
Otherwise if $l\geq \nt-\nr+1$, we have
\begin{IEEEeqnarray}{rCl}
    d_\textnormal{te}^\textnormal{NE}(r) 
    &=& \sum_{i=1}^k(\nt-\nr+2i-1) + l(\nr-k-r), \ k\in\{ 1,\cdots, \nr-r\}, \label{eq:dte2}
\end{IEEEeqnarray}
where the choice of $k$ depends on the channel block length $l$ (see \eqref{eq1: optimal ate}, \eqref{eq2: optimal ate}, and \eqref{eq3: optimal ate}).

Last, we characterize $d_\textnormal{NE}^\star(r)$. If $l\geq\nt-\nr+1$ and at any $k$ in set $\{ 1,\cdots, \nr-r\}$, 
\begin{IEEEeqnarray}{rCl}
    d_\textnormal{te}^\textnormal{NE}(r) 
    &\geq& \sum_{i=1}^k(\nt-\nr+2-1) + (\nt-\nr+1)(\nr-k-r)\nonumber\\
    &=&(\nt-\nr+1)(\nr-r) \nonumber \\
    &=& d_\textnormal{out}^\textnormal{NE}(r).
    \label{eq:dgeqd}
\end{IEEEeqnarray}
The proof in this case is concluded by combining~\eqref{e6-1} with~\eqref{dout negative exponential} and~\eqref{eq:dgeqd}.

 If $l<\nt-\nr+1$, by~\eqref{eq:lleqlth} and~\eqref{dout negative exponential}, we have 
\begin{IEEEeqnarray}{rCl}
    d_\textnormal{te}^\textnormal{NE}(r) 
    &<&(\nt-\nr+1)(\nr-r) \nonumber\\
    &=& d_\textnormal{out}^\textnormal{NE}(r).
    \label{eq:ll}
\end{IEEEeqnarray}
Combing \eqref{e6-2} with~\eqref{dout negative exponential},~\eqref{eq:dte} and~\eqref{eq:ll}, the proof is concluded in this case.

\end{IEEEproof}
\medskip

\begin{remark}
Notice that in \eqref{dout bound for negative exponential},
the difference in the upper and lower bounds on $d_\textnormal{out}^\textnormal{NE}(r)$ lies in the optimization domain. It is easily verified that set $\set{A}$ represents a larger set compared with the original outage set $\set{C}$, while set $\set{B}$ is a smaller set, i.e.,
 \begin{align}
    \set{B} \subseteq \set{C} \subseteq \set{A}. \label{relation of three sets}
 \end{align}
The results above show that the infimums over different optimization domains are achieved at the same point, revealing that the bounds derived in Proposition~\ref{proposition: mutual information bounds} are sufficiently tight in terms of diversity gain.
\end{remark}
\begin{remark}
Observing \eqref{eq1: optimal ate} and \eqref{eq3: optimal ate}, we can find that $a_\textnormal{te}^\star$ has at most $(\nr-r)$ entries equal to $2$. Intuitively, note that $(\set{B}^c)^\prime$ denotes the feasible regime in \eqref{new expression of dG for ne channel}, thus any point $a$ in it satisfies $2\nr-\sum_{i=1}^{\nr}a_{i}\geq 2r$. This indicates that any $a$ has at most $(\nr-r)$ entries equal to 2.
\end{remark}
\section{DMT Analysis on Gamma-Gamma Channel}
\label{sec:Gamma-Gamma Atmospheric Turbulence}
In this section, we consider the OWC channel with atmospheric turbulence according to gamma-gamma distribution. This distribution is commonly used to model the channel with moderate-to-strong atmospheric turbulence intensity. The outage diversity and the optimal DMT are characterized in the following.
\subsection{Outage Diversity Gain Characterization} 
Due to the presence of $K_{\tau}(\cdot)$ in \eqref{gamma gamma pdf}, it is difficult to analyze original gamma-gamma distribution directly. Instead, we choose an alternative expression in terms of power series, i.e.,
\begin{IEEEeqnarray}{rCl}
        {f}({h}_{i j}^{r})=\sum_{n=0}^{\infty} d_{n}(\rho_{1}, \rho_{2})({h}_{i j}^{r})^{n+\rho_{2}-1}+\sum_{n=0}^{\infty} d_{n}(\rho_{2},\rho_{1})({h}_{i j}^{r})^{n+\rho_{1}-1},
    \end{IEEEeqnarray}
where 
\begin{IEEEeqnarray}{rCl}
    d_{n}(\rho_{1}, \rho_{2})=\frac{(\rho_{1}\rho_{2})^{\rho_{2}+n}\Gamma(\rho_{1}-\rho_{2})}{\Gamma(\rho_{1})\Gamma(\rho_{2})(1+\rho_{2}-\rho_{1})_nn!}.
\end{IEEEeqnarray}
Combined with \eqref{h_ij component} and \eqref{MIMO p(H)}, we obtain the distribution of gamma-gamma channel as
\begin{IEEEeqnarray}{rCl}
    f(\matt{H}) 
    = \prod_{i=1}^{\nr} \prod_{j=1}^{\nt}\Biggl\{&&\sum_{n=0}^{\infty} d_{n}(\rho_{1}, \rho_{2})\left(e^{\nu d_{i j}} h_{i j}\right)^{n+\rho_{2}-1} e^{\nu d_{i j}}\nonumber\\
    && +\sum_{n=0}^{\infty} d_{n}(\rho_{2}, \rho_{1})\left(e^{\nu d_{i j}} h_{i j}\right)^{n+\rho_{1}-1} e^{\nu d_{i j}}\Biggr\}. \label{Gamma-Gamma pH(a)}
\end{IEEEeqnarray}
After some algebraic manipulations, we can rewrite \eqref{Gamma-Gamma pH(a)} as
\begin{IEEEeqnarray}{rCl}
    f(\matt{H}) 
    &\doteq& \prod_{i=1}^{\nr} \prod_{j=1}^{\nt}\Biggl\{\sum_{n=0}^{\infty} d_{n}(\rho_{1}, \rho_{2})\left(e^{\nu d_{i j}}\right)^{n+\rho_{2}}
    (\OSNR)^{-\frac{n+\rho_{2}-1}{2}a_{\nr}} \nonumber\\
    &&\qquad\qquad\quad+\sum_{n=0}^{\infty} d_{n}(\rho_{2}, \rho_{1})\left(e^{\nu d_{i j}}\right)^{n+\rho_{1}}
    (\OSNR)^{-\frac{n+\rho_{1}-1}{2}a_{\nr}} \Biggr\} \\
    &\doteq& (\OSNR)^{-\frac{ \nr\nt a_{\nr}}{2}(\rho-1)}, \label{eq:mmm}
\end{IEEEeqnarray}
where $\rho$ denotes $\min\{\rho_{1},\rho_{2}\}$.

Substituting~\eqref{eq:mmm} into~\eqref{f(a) definition} and \eqref{integration with A}, the outage diversity order $d_\textnormal{out}^\textnormal{GG}(r)$ at multiplexing order $r$ is bounded by
\begin{IEEEeqnarray}{rCl}
    \inf_{a \in \set{A}^{\prime}} d_\textnormal{out}^\textnormal{GG}(r,a) \label{dout bound for gamma-gamma} \leq d_\textnormal{out}^\textnormal{GG}(r) 
    \leq  \inf_{a \in \set{B}^{\prime}} d_\textnormal{out}^\textnormal{GG}(r,a), \label{dout bound for gg channel}
\end{IEEEeqnarray}
where function $d_\textnormal{out}^\textnormal{GG}(r,a)$ is defined as
\begin{IEEEeqnarray}{rCl}
    d_\textnormal{out}^\textnormal{GG}(r,a)=\sum_{i=1}^{\nr} \frac{\left(\nt-\nr+2 i-1\right)}{2} a_{i} +\frac{\nr \nt a_{\nr}}{2}\left(\rho-1\right);
    \label{eq:doutgamma}
\end{IEEEeqnarray}
and where sets $\set{A}^{\prime}$ and $\set{B}^{\prime}$ are defined as in \eqref{set A prime} and \eqref{set B prime}, respectively.
\subsection{Optimal DMT Characterization}
We present the main results in the following theorem.
\begin{theorem}[DMT of Gamma-Gamma Channel]
\label{theorem: gamma-gamma DMT}
Given a channel with distribution in \eqref{Gamma-Gamma pH(a)}, when $\nr=1$, if $l\geq l_\textnormal{th}^\textnormal{GG}=\rho\nt$, the optimal diversity order is 
\begin{IEEEeqnarray}{rCl}
    d_\textnormal{GG}^\star(r) = \rho\nt(1-r),
\end{IEEEeqnarray}
otherwise,
\begin{IEEEeqnarray}{L}
 l(1-r)\leq d_\textnormal{GG}^\star(r) \leq \rho\nt(1-r);
\end{IEEEeqnarray}
when $\nr>1$, if $l\geq l_\textnormal{th}^\textnormal{GG}=\nt-\nr+1$, then
\begin{IEEEeqnarray}{rCl}
    d_\textnormal{GG}^\star(r) = (\nt-\nr+1)(\nr-r),
\end{IEEEeqnarray}
otherwise,
\begin{IEEEeqnarray}{L}
 l(\nr-r)\leq d_\textnormal{GG}^\star(r) \leq (\nt-\nr+1)(\nr-r).
\end{IEEEeqnarray}
\end{theorem}

\begin{IEEEproof}
We follow the similar arguments as in the proof of Theorem~\ref{theorem: negative exponential DMT}. Here we mainly emphasize the differences. 

Compared with the function in \eqref{dout_ne(r,a)}, $d_\textnormal{out}^\textnormal{GG}(r,a)$ in~\eqref{eq:doutgamma} contains an extra term $\nr\nt(\rho-1)a_{\nr}/2$. We consider the cases $\nr=1$ and $\nr>1$ separately to show the differences when $a_{1} = a_{\nr}$ and $a_{1} \neq a_{\nr}$.  

When $\nr=1$, the infimums in~\eqref{dout bound for gg channel} are both achieved at point $a_\textnormal{out}^\star=2(\nr-r)$. Hence, the outage diversity gain is given by
\begin{IEEEeqnarray}{rCl}
    d_\textnormal{out}^\textnormal{GG}(r)
    &=&\left(\nt - \nr + 1 +\nr\nt(\rho-1)\right)(\nr-r)\nonumber\\
    &=&\rho\nt(1-r).\label{eq1: dout for gg channel}
\end{IEEEeqnarray}

When $\nr>1$, the infimums in~\eqref{dout bound for gg channel} are both achieved at point $a_\textnormal{out}^\star=[2(\nr-r),0,\ldots,0]$, and we have
\begin{IEEEeqnarray}{rCl}
    d_\textnormal{out}^\textnormal{GG}(r)=(\nt-\nr+1)(\nr-r).\label{eq2: dout for gg channel}
\end{IEEEeqnarray}

Now we calculate $d_\textnormal{te}^\textnormal{GG}(r)$. Modifying the optimization domain in \eqref{d_G definition} with $\set{A}^c$ or $\set{B}^c$, we have 
\begin{IEEEeqnarray}{rCl}
    \inf_{(\set{B}^c)^{\prime}} d_\textnormal{te}^\textnormal{GG}(r,a) 
    \leq  
    d_\textnormal{te}^\textnormal{GG}(r)
    \leq \inf_{(\set{A}^c)^{\prime}} d_\textnormal{te}^\textnormal{GG}(r,a),\label{dG bound for gg channel}
\end{IEEEeqnarray}
where function $d_\textnormal{te}^\textnormal{GG}(r,a)$ is defined as
\begin{IEEEeqnarray}{rCl}
    d_\textnormal{te}^\textnormal{GG}(r,a)
    =\sum_{i=1}^{\nr} \frac{\nt-\nr+2i-1}{2} a_i +\frac{\nr \nt a_{\nr} }{2} \left(\rho -1 \right)  +\frac{l}{2} \left( \sum_{i=1}^{\nr} (2-a_i)^+-2r \right);\nonumber\\
	\label{eq:gra}
\end{IEEEeqnarray}
and where sets $(\set{A}^c)^{\prime}$ and $(\set{B}^c)^{\prime}$ are defined as in \eqref{Ac prime} and \eqref{Bc prime}, respectively.

Compared with the function in \eqref{dte(r,a) for ne channel}, $d_\textnormal{te}^\textnormal{GG}(r,a)$ also has an extra term $\nr\nt(\rho-1)a_{\nr}/2$. We still separately consider the cases when $\nr=1$ and $\nr>1$.  

When $\nr=1$, if $l < \nt - \nr +1 + \nr\nt(\rho-1)=\rho\nt$, the infimums in \eqref{dG bound for gg channel} are both achieved at $a_\textnormal{te}^\star=0$, which yields
\begin{IEEEeqnarray}{rCl}
    d_\textnormal{te}^\textnormal{GG}(r)=l(\nr-r),
\end{IEEEeqnarray} 
otherwise the infimums in \eqref{dG bound for gg channel} are both achieved at $a_\textnormal{te}^\star=2(\nr-r)$, thus 
\begin{IEEEeqnarray}{rCl}
    d_\textnormal{te}^\textnormal{GG}(r)
    &=&(\nt-\nr+1+\nr\nt(\rho-1))(\nr-r)\nonumber\\
    &=&\rho\nt(1-r).
\end{IEEEeqnarray}

When $\nr>1$, since sets $(\set{A})^\prime$ and $(\set{B})^\prime$ in \eqref{dG bound for gg channel} are the same as the corresponding sets in the negative exponential channel in \eqref{new expression of dG for ne channel}, here we just need to show $d_\textnormal{te}^\textnormal{GG}(r)= d_\textnormal{te}^\textnormal{NE}(r)$. On the one hand, as $\nr\nt(\rho-1)a_{\nr}/2\geq0$ for any $a_{\nr}\geq0$, we can obtain that $d_\textnormal{te}^\textnormal{GG}(r,a)\geq d_\textnormal{te}^\textnormal{NE}(r,a)$ at any $a$ with $a_{\nr}\geq0$. On the other hand, at the point $a_\textnormal{te}^\star$ defined in \eqref{eq1: optimal ate}, \eqref{eq2: optimal ate}, and \eqref{eq3: optimal ate}, we have $d_\textnormal{te}^\textnormal{GG}(r,a_\textnormal{te}^\star)=d_\textnormal{te}^\textnormal{NE}(r,a_\textnormal{te}^\star)$. The proof is concluded.

\end{IEEEproof}
\medskip

\begin{remark}
Notice that when $\nr>1$, the $\nr$th entry of $a_\textnormal{out}^\star$ or $a_\textnormal{te}^\star$ is always $0$, which indicates the extra term $\nr\nt(\rho-1)a_{\nr}/2$ plays no role on $d_\textnormal{out}^\textnormal{GG}(r,a)$ and $d_\textnormal{te}^\textnormal{GG}(r,a)$. Hence when $\nr>1$, we always have $d_\textnormal{out}^\textnormal{GG}(r)=d_\textnormal{out}^\textnormal{NE}(r)$ and $d_\textnormal{te}^\textnormal{GG}(r)=d_\textnormal{te}^\textnormal{NE}(r)$. 
\end{remark}

\section{DMT Analysis on Log-Normal Channel}
In this section, we consider the OWC channel with atmospheric turbulence according to log-normal distribution. This distribution is commonly used to model the channel with weak atmospheric turbulence intensity. The outage diversity order and the optimal DMT are characterized in the following.
\label{sec:Log-Normal Atmospheric Turbulence}
\subsection{Outage Diversity Gain Characterization} 
Substituting \eqref{h_ij component} and \eqref{log normal pdf} into \eqref{MIMO p(H)}, we have
\begin{IEEEeqnarray}{rCl}
    f(\matt{H}) 
    &=&\prod_{i=1}^{\nr} \prod_{j=1}^{\nt} \left\{\frac{1}{ h_{i j} \sqrt{2 \pi {\sigma}_{l}^{2}}} \exp \left(-\frac{\Bigl(\log\left(\exp \left(\nu d_{i j}\right) h_{i j}\right)-\mu_{l}\Bigr)^{2}}{2 {\sigma}_{l}^{2}}\right) \right\}.\label{whole pdf log-normal}
\end{IEEEeqnarray}
Following similar arguments as in Section~\ref{subsec-outage diversity gain characterization}, we simplify $f(\matt{H})$ as
\begin{IEEEeqnarray}{rCl}
   f(\matt{H})
    &\doteq& (\OSNR)^{\frac{\nr\nt a_{\nr}}{2}} \times \prod_{i=1}^{\nr} \prod_{j=1}^{\nt} \left\{\exp \left(-\frac{\biggl(\log\Bigl(\exp \left(\nu d_{i j}\right) (\OSNR)^{-\frac{a_{\nr}}{2}}\Bigr)-\mu_{l}\biggr)^{2}}{2 {\sigma}_{l}^{2}}\right)\right\}\quad\nonumber\\
    &=& (\OSNR)^{\frac{\nr\nt a_{\nr}}{2}} \times \prod_{i=1}^{\nr} \prod_{j=1}^{\nt} \left\{\exp \left(-\frac{\Bigl(\log\bigl(\OSNR\bigr)^{-\frac{a_{\nr}}{2}}-\bigl(\mu_{l}-\nu d_{i j}\bigr)\Bigr)^{2}}{2 {\sigma}_{l}^{2}}\right)\right\}\quad\nonumber\\   
    &=& (\OSNR)^{\frac{\nr\nt a_{\nr}}{2}} \times  \exp \Biggl(-\log(\OSNR)\times\frac{\nr\nt}{2\sigma_{l}^2}\left(\frac{a_{\nr}^{2}}{4} \log(\OSNR)+a_{\nr} \beta_1 + \frac{\beta_2}{\log(\OSNR)}\right)\Biggr)\quad\nonumber\\   
    &=& (\OSNR)^{\frac{\nr\nt a_{\nr}}{2}} \times (\OSNR)^{\log_\OSNR \exp \left(-\log(\OSNR)\times\frac{\nr\nt}{2\sigma_{l}^2}\Bigl(\frac{a_{\nr}^{2}}{4} \log(\OSNR)+a_{\nr} \beta_1 + \frac{\beta_2}{\log(\OSNR)}\Bigr)\right) }\quad\nonumber\\
    &=&(\OSNR)^{\frac{\nr \nt a_{\nr}}{2}-\frac{\nr \nt}{2 {\sigma}_{l}^{2}}\left(\frac{a_{\nr}^{2}}{4} \log(\OSNR)+a_{\nr} \beta_1 + \frac{\beta_2}{\log(\OSNR)}\right)}\label{long ep}\\
    &\doteq&(\OSNR)^{\frac{\nr \nt a_{\nr}}{2}-\frac{\nr \nt}{2 {\sigma}_{l}^{2}}\left(\frac{a_{\nr}^{2}}{4} \log(\OSNR)+a_{\nr} \beta_1\right)},\label{eq:logprob}
\end{IEEEeqnarray}
where $\log_\OSNR(\cdot)$ denotes the logarithm to the base of $\OSNR$; where $\beta_1$ and $\beta_2$ are defined as
\begin{align}
    \beta_1 &=\frac{\sum_{i=1}^{\nr}\sum_{j=1}^{\nt}\left(\mu_{l}-\nu d_{ij}\right)}{\nr \nt}, \label{beta_1 definition}\\
    \beta_2 &= \frac{\sum_{i=1}^{\nr}\sum_{j=1}^{\nt} \left(\mu_{l}-\nu d_{i j}\right)^{2}}{\nr \nt}\label{beta_2 definition};
\end{align}
and where \eqref{long ep} follows by
\begin{IEEEeqnarray}{rCl}
    (\OSNR)^{\log_{\OSNR}e^{b\cdot\log(\OSNR)}}
    &=& (\OSNR)^{b\cdot \log_{\OSNR}e^{\log(\OSNR)}} \\
    &=&(\OSNR)^b.
\end{IEEEeqnarray}
Substituting~\eqref{eq:logprob} into~\eqref{f(a) definition} and \eqref{integration with A}, we bound the outage diversity gain as 
\begin{align}
  \inf_{a \in \set{A}^{\prime}} d_\textnormal{out}^\textnormal{LN}(r,a)  \leq d_\textnormal{out}^\textnormal{LN}(r) \leq \inf_{a \in \set{B}^{\prime}} d_\textnormal{out}^\textnormal{LN}(r,a),  \label{dout bound for log-normal}
\end{align}
where function $d_\textnormal{out}^\textnormal{LN}(r,a)$ is defined as
\begin{align}
   d_\textnormal{out}^\textnormal{LN}(r,a) = &\sum_{i=1}^{\nr} \frac{\left(\nt-\nr+2 i-1\right)}{2} a_{i}-\frac{\nr \nt a_{\nr}}{2} + \frac{\nr \nt}{2{\sigma}_{l}^{2}}\left(\frac{a_{\nr}^{2}}{4} \log(\OSNR)+a_{\nr} \beta_1 \right); \label{dout(r,a) for ln channel}
\end{align}
and where sets $\set{A}^{\prime}$ and $\set{B}^{\prime}$ are defined as in \eqref{set A prime} and \eqref{set B prime}, respectively.
\subsection{Optimal DMT Characterization}
We present the main results in the following theorem.
\begin{theorem}[DMT of Log-Normal Channel]
\label{theorem: log-normal DMT}
Given a channel with distribution in \eqref{whole pdf log-normal}, and denote $\theta \eqdef \Bigl(\log(\OSNR)/2+\beta_1\Bigr)/{\sigma}_{l}^{2}$. When $\nr =1$, if $l\geq l_\textnormal{th}^\textnormal{LN}=  \theta\nt$, the optimal diversity order is
\begin{IEEEeqnarray}{rCl}
    d_\textnormal{LN}^\star(r) = \theta\nt(1-r),\label{exact diversity order for ln channel}
\end{IEEEeqnarray}
otherwise, 
\begin{IEEEeqnarray}{rCl}
  l(1-r)\leq d_\textnormal{LN}^\star(r) \leq \theta\nt(1-r);
  \label{eq:dln}
\end{IEEEeqnarray}
when $\nr>1$, if $l\geq l_\textnormal{th}^\textnormal{LN}=\nt-\nr+1$, then
\begin{IEEEeqnarray}{rCl}
    d_\textnormal{LN}^\star(r) = (\nt-\nr+1)(\nr-r),
    \label{eq:dln2}
\end{IEEEeqnarray}
otherwise,
\begin{IEEEeqnarray}{L}
 l(\nr-r)\leq d_\textnormal{LN}^\star(r) \leq (\nt-\nr+1)(\nr-r).
\end{IEEEeqnarray}

\end{theorem}

\begin{IEEEproof}
We first calculate $d_\textnormal{out}^\textnormal{LN}(r)$. In \eqref{dout(r,a) for ln channel}, $d_\textnormal{out}^\textnormal{LN}(r,a)$ is a nonlinear function to vector $a$ due to the extra term
\begin{IEEEeqnarray}{rCl}
    u(a_{\nr}) \eqdef \frac{\nr\nt}{2 {\sigma}_l^2}\left(\frac{a_{\nr}^2}{4}\log(\OSNR) +a_{\nr} \beta_1  \right).\label{eq1: extra term for ln channel}
\end{IEEEeqnarray}
Notice that at high $\OSNR$, $u(a_{\nr})$ tends to be infinite, and dominates the sum of rest terms. For this reason, we prioritize small $a_{\nr}$ to optimize $d_\textnormal{out}^\textnormal{LN}(r,a)$ over $\set{A}^\prime$ and $\set{B}^\prime$ in \eqref{dout bound for log-normal}. Since this extra term $u(a_{\nr})$ only relates to $a_{\nr}$, we still separately consider $\nr=1$ and $\nr>1$ to distinguish $a_{1}=a_{\nr}$ and $a_{1}\neq a_{\nr}$.

When $\nr=1$, the integral domains $\set{A}^\prime$ and $\set{B}^\prime$ in~\eqref{integration with A} coincide as $\{a\geq 2(\nr-r)\}$. We can obtain the optimal point $a_\textnormal{out}^\star =2(\nr-r)$, and we have 
\begin{IEEEeqnarray}{rCl}
   d_\textnormal{out}^\textnormal{LN}(r)
   &=&\left(\nt-\nr+1-\nr \nt+\frac{\nr \nt}{2 {\sigma}_{l}^{2}}\Bigl(\log(\OSNR)+2 \beta_1 \Bigr)\right)(\nr-r)\nonumber\\
   &=&\theta\nt(1-r).\label{eq1: dout for ln channel}
\end{IEEEeqnarray}

When $\nr>1$, a necessary condition to have finite infimums in~\eqref{dout bound for log-normal} is the $\nr$th entry of $a$ to be zero, i.e., $a_{\nr}=0$. Then the rest of entries in optimal point $a_\textnormal{out}^\star$ can be computed by linear optimization, and we obtain $a_\textnormal{out}^\star =[2(\nr-r),0,\cdots,0]$. Substituting $a_\textnormal{out}^\star$ into \eqref{dout bound for log-normal}, the outage diversity gain is given by
\begin{IEEEeqnarray}{rCl}
    d_\textnormal{out}^\textnormal{LN}(r)=(\nt-\nr+1)(\nr-r).\label{eq2: dout for ln channel}
\end{IEEEeqnarray}

 Now we calculate $d_\textnormal{te}^\textnormal{LN}(r)$. Modifying the integral domain in \eqref{d_G definition} as $\set{A}^c$ or $\set{B}^c$, we have 
 \begin{IEEEeqnarray}{rCl}
    \inf_{(\set{B}^c)^{\prime}} d_\textnormal{te}^\textnormal{LN}(r,a) \leq d_\textnormal{te}^\textnormal{LN}(r)
    \leq\inf_{(\set{A}^c)^{\prime}} d_\textnormal{te}^\textnormal{LN}(r,a),\label{dG bound for ln channel}
\end{IEEEeqnarray}
where function $d_\textnormal{te}^\textnormal{LN}(r,a)$ is defined as
\begin{IEEEeqnarray}{rCl}
    d_\textnormal{te}^\textnormal{LN}(r,a)
    &=& \sum_{i=1}^{\nr} \frac{\nt-\nr+2i-1}{2} a_i 
	-\frac{\nr \nt}{2}a_{\nr} \notag \\
	&&  + \frac{\nr\nt}{2 {\sigma}_l^2}\biggl(\frac{a_{\nr}^2}{4}\log(\OSNR) +a_{\nr} \beta_1 \biggr) + \frac{l}{2} \biggl(\sum_{i=1}^{\nr} (2-a_i)^+-2r \biggr);
\end{IEEEeqnarray}
and where sets $(\set{A}^c)^{\prime}$ and $(\set{B}^c)^{\prime}$ are defined as in \eqref{Ac prime} and \eqref{Bc prime}, respectively.

It is direct to see that $d_\textnormal{te}^\textnormal{LN}(r,a)$ is also a nonlinear function due to the extra term $u(a_{\nr})$. Using similar method argued above, we can directly show that when $\nr=1$, if $l<\theta \nt$, both infimums of $d_\textnormal{te}^\textnormal{LN}(r,a)$ over $(\set{A}^c)^\prime$ and $(\set{B}^c)^\prime$ are achieved at $a_\textnormal{te}^\star=0$, which yields
\begin{IEEEeqnarray}{rCl}
    d_\textnormal{te}^\textnormal{LN}(r)=l(\nr-r),
\end{IEEEeqnarray}
otherwise, $a_\textnormal{te}^\star=2(\nr-r)$, and we have
\begin{IEEEeqnarray}{rCl}
    d_\textnormal{te}^\textnormal{LN}(r)
    &=&\left(\nt-\nr+1-\nr \nt+\frac{\nr \nt}{2 {\sigma}_{l}^{2}}\Bigl(\log(\OSNR)+2 \beta_1 \Bigr)\right)(\nr-r)\nonumber\\
    &=&\theta\nt(1-r).
\end{IEEEeqnarray}

When $\nr>1$, following similar arguments as in the derivation of $d_\textnormal{out}^\textnormal{LN}(r)$, we obtain the optimal point $a_\textnormal{te}^\star$ has the same expressions as in \eqref{eq1: optimal ate}, \eqref{eq2: optimal ate}, and \eqref{eq3: optimal ate}, and $d_\textnormal{te}^\textnormal{LN}(r)$ has the same expressions as in~\eqref{eq:dte} and~\eqref{eq:dte2}. Then the proof is concluded.

\end{IEEEproof}
\medskip

\begin{remark}
The term $u(a_{\nr})$ causes the non-linearity of $d_\textnormal{out}^\textnormal{LN}(r,a)$ in \eqref{dout(r,a) for ln channel}. When $\nr>1$, a necessary condition to have a finite $d_\textnormal{out}^\textnormal{LN}(r)$ is $a_{\nr}=0$, and hence $u(a_{\nr})=0$. Eliminating the influence of $a_{\nr}$, the optimization problems in \eqref{dout bound for log-normal} are essentially equivalent to the linear problems in \eqref{dout bound for negative exponential}. 
\end{remark}

One issue in Theorem~\ref{theorem: log-normal DMT} is when $\nr=1$, at high $\OSNR$, the optimal diversity gain $d_\textnormal{LN}^\star(r)$ tends to be infinite, which fails to reflect the performance limits of practical OWC systems. In fact, similar phenomena are also observed in the existing literature \cite{Safari2008,Pan2014,Hatef2016}. This issue is due to the fact that $d_\textnormal{LN}^\star(r)$ contains the $\OSNR$ related term, while the diversity gain essentially measures the system performance under high $\OSNR$. To deal with this problem, we adopt the definition of asymptotically relative diversity order (ARDO) proposed in \cite{Safari2008}, and then analyze the optimal diversity gain of log-normal channel. The ARDO of a transmission scheme is given by
\begin{IEEEeqnarray}{rCl}
    \hat{d}^\star(r)
    &=& \lim _{\OSNR \rightarrow \infty} \frac{d^\star(r)}{d^\star_\textnormal{BM}(r)},
\end{IEEEeqnarray}
where $d^\star(r)$ denotes the optimal diversity order, and where $d^\star_\textnormal{BM}(r)$ denotes the optimal diversity order of a benchmark scheme. In the following, we use the outage diversity order at $r=0$ of the SISO system as the benchmark\footnote{The benchmark is chosen by considering the fact that outage diversity order at $r=0$ is the maximal achievable diversity order for any transmission scheme.}, such that
\begin{IEEEeqnarray}{rCl}
    \hat{d}_\textnormal{LN}^\star(r)=\frac{d_{\textnormal{LN}}^\star(r)}{d_\textnormal{out}^\textnormal{LN}(0)}.
    \label{eq:lnardo}
\end{IEEEeqnarray}
Substituting $d_\textnormal{LN}^\star(r)$ in~\eqref{eq:dln} and~\eqref{eq:dln2} into~\eqref{eq:lnardo}, we have the following proposition: 
\begin{proposition}
Given a channel with distribution in \eqref{whole pdf log-normal}, when $\nr=1$, the ARDO under a large block length is given by
\begin{IEEEeqnarray}{rCl}
    \hat{d}_\textnormal{LN}^\star(r) =\nt(1-r);
\end{IEEEeqnarray}
under a small block length we have 
\begin{IEEEeqnarray}{rCl}
  0\leq \hat{d}_\textnormal{LN}^\star(r) \leq \nt(1-r).
\end{IEEEeqnarray}
\end{proposition}



\section{Numerical Results and Discussions}
\label{sec: Numerical Examples}
\subsection{Optimal DMT Analysis}
In this section, we present numerical examples of the derived DMT results. For the notational convenience, a channel with $\nr$ receive, $\nt$ transmit antennas, and $l$ block length is simply denoted by $\nt\times\nr\times l$. 

\begin{figure}
  \begin{center}
  \includegraphics[width=3.5in]{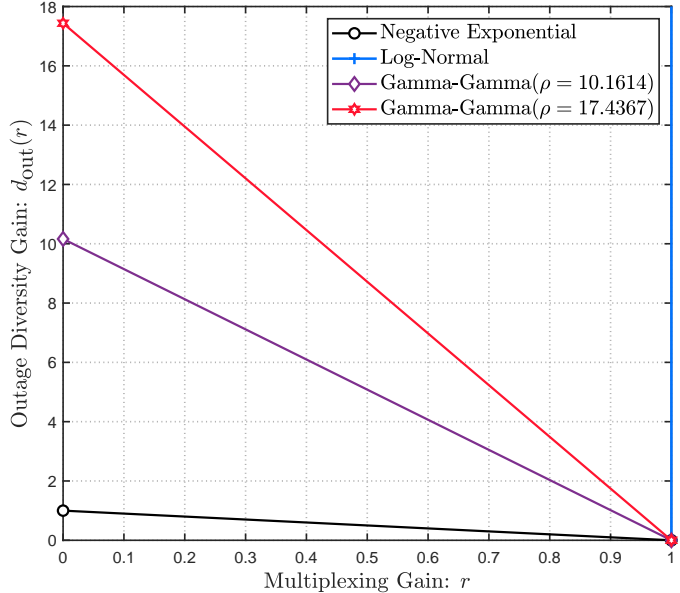}\\
  \caption{Outage diversity gain for SISO channel.}\label{figure:siso-dout}
  \end{center}
\end{figure}
\begin{figure}
  \begin{center}
  \includegraphics[width=3.5in]{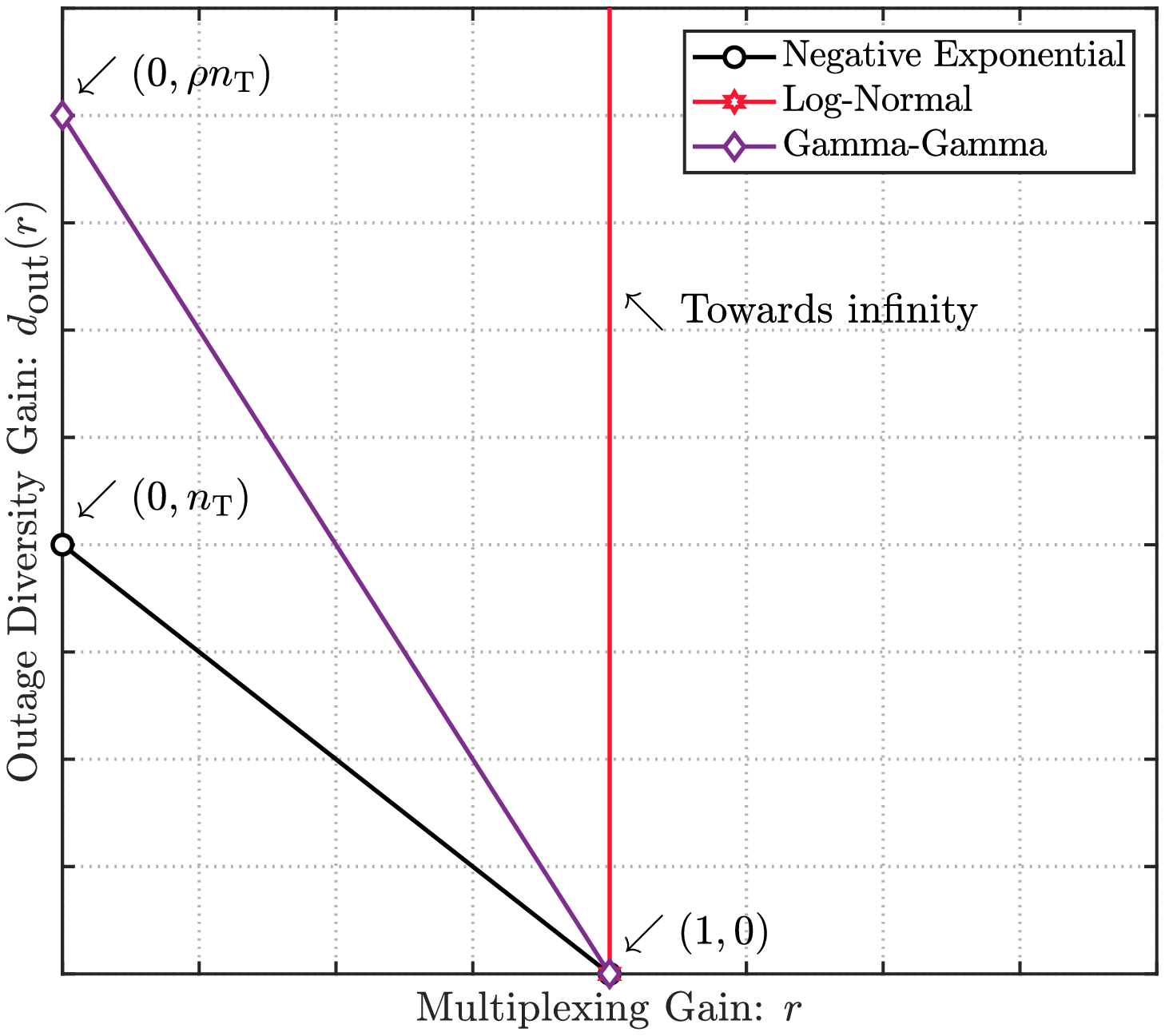}\\
  \caption{Outage diversity gain for $\nt\times1$ MISO channel.}\label{figure:miso-dout}
  \end{center}
\end{figure}
\begin{figure}
  \begin{center}
  \includegraphics[width=3.5in]{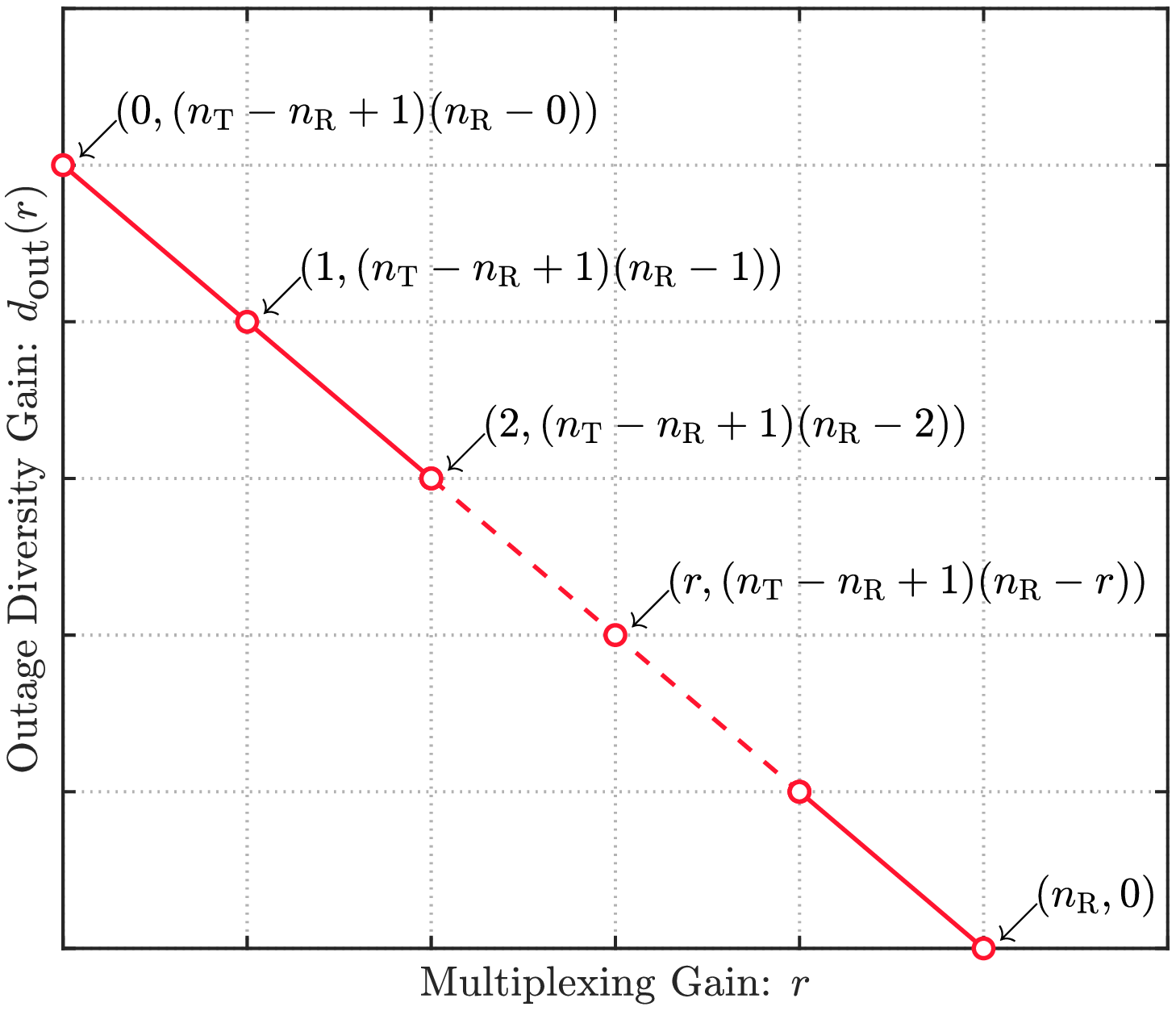}\\
  \caption{Outage diversity gain for $\nt\times\nr$ MIMO channel.}\label{figure:mimo-dout}
  \end{center}
\end{figure}

Figures~\ref{figure:siso-dout}, \ref{figure:miso-dout}, and \ref{figure:mimo-dout} depict the outage diversity gain $d_\textnormal{out}(r)$ at the multiplexing gain $r$ under different channel conditions for SISO, MISO, and MIMO channels. These curves can also represent the optimal spatial diversity gain $d^\star(r)$ when $l\geq l_\textnormal{th}$, where $l_\textnormal{th}$ represents the block length threshold of corresponding channels, such as $l_\textnormal{th}^\textnormal{NE}$, $l_\textnormal{th}^\textnormal{GG}$, $l_\textnormal{th}^\textnormal{LN}$ (see Theorems~\ref{theorem: negative exponential DMT}, \ref{theorem: gamma-gamma DMT}, and \ref{theorem: log-normal DMT}). 

Figure~\ref{figure:siso-dout} shows the derived outage diversity gain $d_\textnormal{out}(r)$ under three different channel distributions, i.e., negative exponential, gamma-gamma, and log-normal channels. Besides, we choose two different values of $\rho$ given in \cite{Parikh2011} to show their influence on the gamma-gamma channel, i.e., $(1)\ \rho= 10.1614$, $(2)\ \rho=17.4369$. For negative exponential and gamma-gamma channels, $d_\textnormal{out}(r)$ linearly decreases with the increasing $r$. For the log-normal channel, $d_\textnormal{out}(r)$ tends to be infinite when $r=0$ and returns to $0$ when $r$ takes the maximum value of $1$. Since log-normal, gamma-gamma, and negative exponential distributions model the channel with weak, moderate, and strong atmospheric turbulence intensities, respectively, this figure also reflects the fact that increasing turbulence intensity may hamper the diversity gain. As shown in Figure~\ref{figure:miso-dout}, the above conclusions also hold in the MISO channels. 

From Figure~\ref{figure:mimo-dout}, we observe that the maximum outage and multiplexing gains are given by $d^\star_\textnormal{max} = (\nt-\nr+1)\nr$ and $r^\star_\textnormal{max} = \nr$, and the outage diversity gains are equal under these three different fading MIMO channels. 
\begin{remark}
Although different distributions may lead to different outage probabilities, the result in Figure~\ref{figure:mimo-dout} reveals that the exponents of outage probabilities are identical regardless of the fading properties. In fact, this phenomenon has been explained in \cite[Theorem~21]{Leizhao2007}. Specifically, as long as some relevant attributes by different channels are identical, the derived outage diversity gain will be the same. Therefore, we could infer that the result shown in Figure~\ref{figure:mimo-dout} is also caused by some internal attributes which behave the same among these three fading channels.
\end{remark}

\begin{figure}
  \begin{center}
  \includegraphics[width=3.5in]{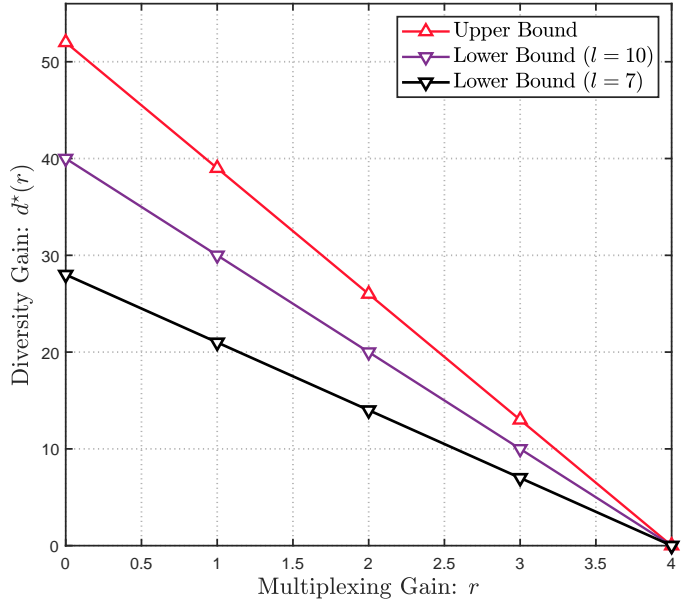}\\
  \caption{Bounds on optimal diversity gain for $16\times4$ MIMO channel.}\label{figure: d when l smaller that lth}
  \end{center}
\end{figure}

When $l < l_\textnormal{th}$, Figure~\ref{figure: d when l smaller that lth} depicts the bounds on optimal $d^{\star}(r)$ for a $16\times4$ MIMO channel with $l  =7$, and $l =10$, respectively. Here, $l_\textnormal{th} =13$. It is direct to see $d^{\star}(r)$ is upper-bounded by $d_\textnormal{out}(r)$ and lower-bounded by $d_\textnormal{te}(r)$ for the above channels. For any $l<l_\textnormal{th}$ and $r< \nr$, the lower bound on diversity gain is strictly below the upper bound, implying that there is always uncertainty on the characterization of optimal DMT curve. The uncertainty tends to decrease as the block length or the multiplexing gain increases.

\begin{figure}
  \begin{center}
  \includegraphics[width=3.5in]{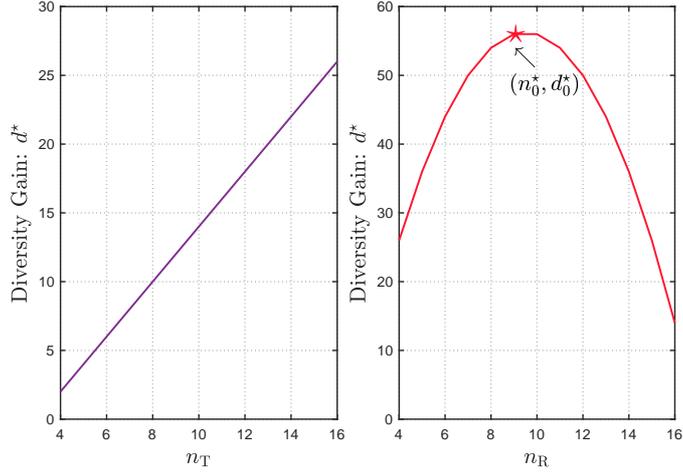}\\
  \caption{Optimal diversity gain at fixed $\nr$ or $\nt$.}\label{figure: d relation with nT and nR}
  \end{center}
\end{figure}

We also depict the optimal $d^{\star}(r)$ as a function of the number of transmit antenna $\nt$ or receive antenna $\nr$ in Figure~\ref{figure: d relation with nT and nR}. The left curve of Figure~\ref{figure: d relation with nT and nR} depicts the optimal diversity gain $d^{\star}(r)$ as a function of $\nt$ with fixed parameters $\nr=4$ and $r=2$, while the right curve depicts $d^{\star}(r)$ as a function of $\nr$ with fixed parameters $\nt=16$ and $r=2$. We observe that $d^{\star}(r)$ monotonically increases as $\nt$ increases, while increasing $\nr$ does not always guarantee larger diversity gain. In fact, our derived $d^{\star}(r)$ is a linear function with $\nt$, but a second-order function of $\nr$, which is fundamentally different from the RF DMT result established before \cite{Lizhong_Zheng2003}.

\begin{remark}
Since the optimal diversity order $d^\star(r)$ is a second-order function of $\nr$, hence, installing ${\nr}^\star=\lfloor \frac{\nt+1+r}{2} \rfloor$ antennas at the receiving end may achieve the best performance.
\end{remark}

\subsection{DMT Comparison with Existing Results}
\begin{table*}\scriptsize 
\caption{Results of different MIMO channels}
\begin{center}
\begin{tabular}{|c|c|c|c|}
\hline
&$d_\textnormal{out}(r)$&$d_\textnormal{te}(r)\ (l\leq l_\textnormal{th})$&$l_\textnormal{th}$\\ 
\hline
Rayleigh~\cite{Lizhong_Zheng2003}&$(\nt-r)(\nr-r)$&$-l(r-r_{1})+(\nt-r_{1})(\nr-r_{1})$&$\nt+\nr-1$\\
\hline
Negative Exponential~\cite{Jaiswal2019}&$\frac{1}{2}(\nt-r)(\nr-r)$&$-\frac{l}{2}(r-r_{1}) +\frac{1}{2} (\nt-r_{1})(\nr-r_{1})$&$\nt+\nr-1$\\
\hline
Negative Exponential (us)&$(\nt-\nr+1)(\nr-r)$&$l(\nr-r)$&$\nt-\nr+1$\\
\hline
\end{tabular}
\label{tab1}
\begin{tablenotes}\footnotesize
\item[1] where $r_{1}=\nr- \lceil{(l-|\nt-\nr|-1)}/{2} \rceil$. 
\end{tablenotes}
\end{center}
\end{table*}

Table~\ref{tab1} compares our derived DMT results with existing results in RF and OWC channels. Note that the slope of $d_\textnormal{out}(r)$ or $d_\textnormal{te}(r)$ in~\cite{Lizhong_Zheng2003} is two times large as in~\cite{Jaiswal2019}, and the threshold of block length in~\cite{Lizhong_Zheng2003} is the same as in~\cite{Jaiswal2019}, which are different from our derived results. In the following, to avoid ambiguity, we denote $l_{\textnormal{th}}$ in~\cite{Lizhong_Zheng2003} or~\cite{Jaiswal2019} as $l^{\textnormal{ZJ}}_{\textnormal{th}}$.

\begin{figure}
  \begin{center}
  \includegraphics[width=3.5in]{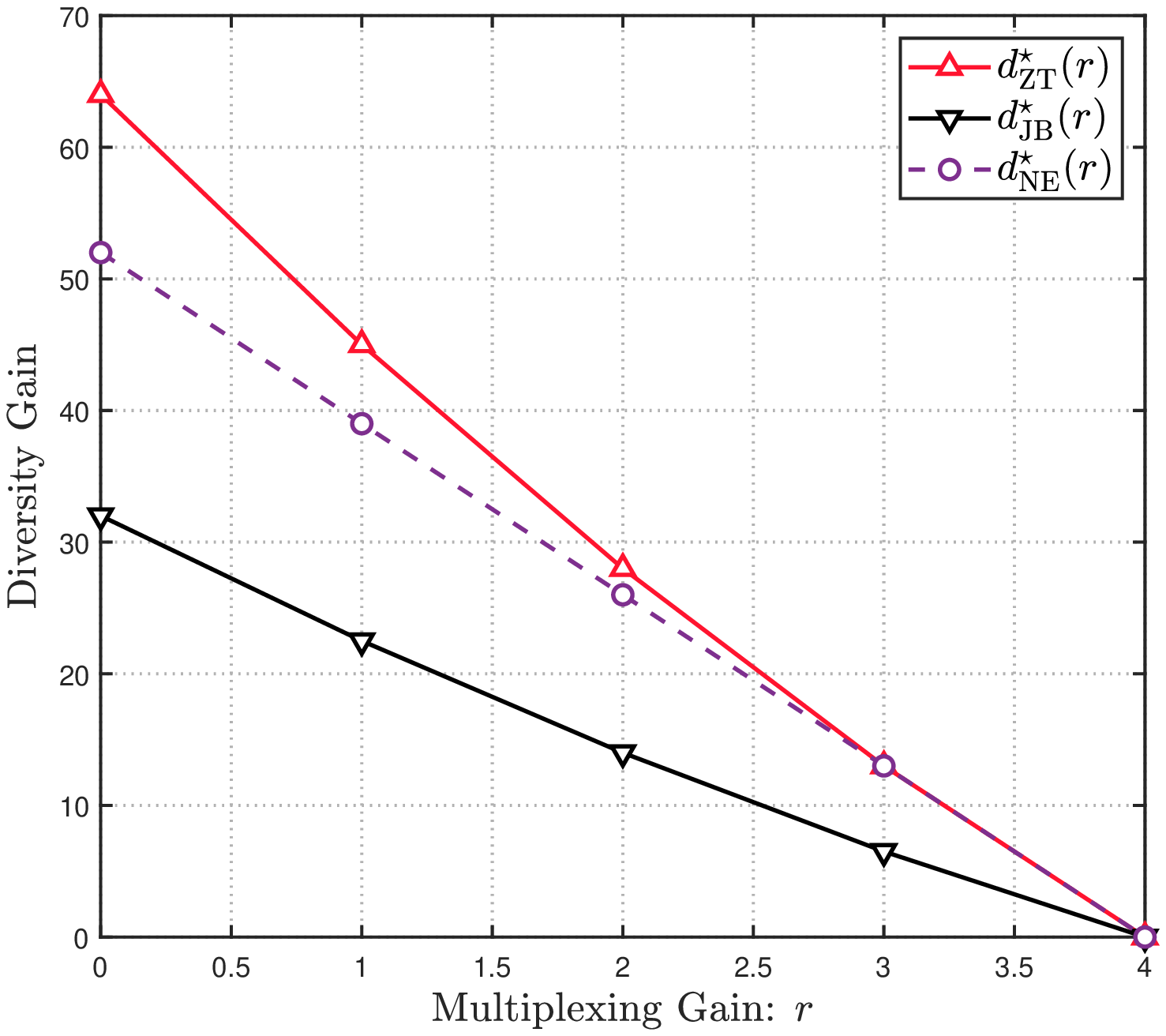}\\
  \caption{Comparisons of the optimal DMT curve for $16\times4\times\infty$ MIMO channel.}\label{figure: d-compare-16-4}
  \end{center}
\end{figure}

\begin{figure}
  \begin{center}
  \includegraphics[width=3.5in]{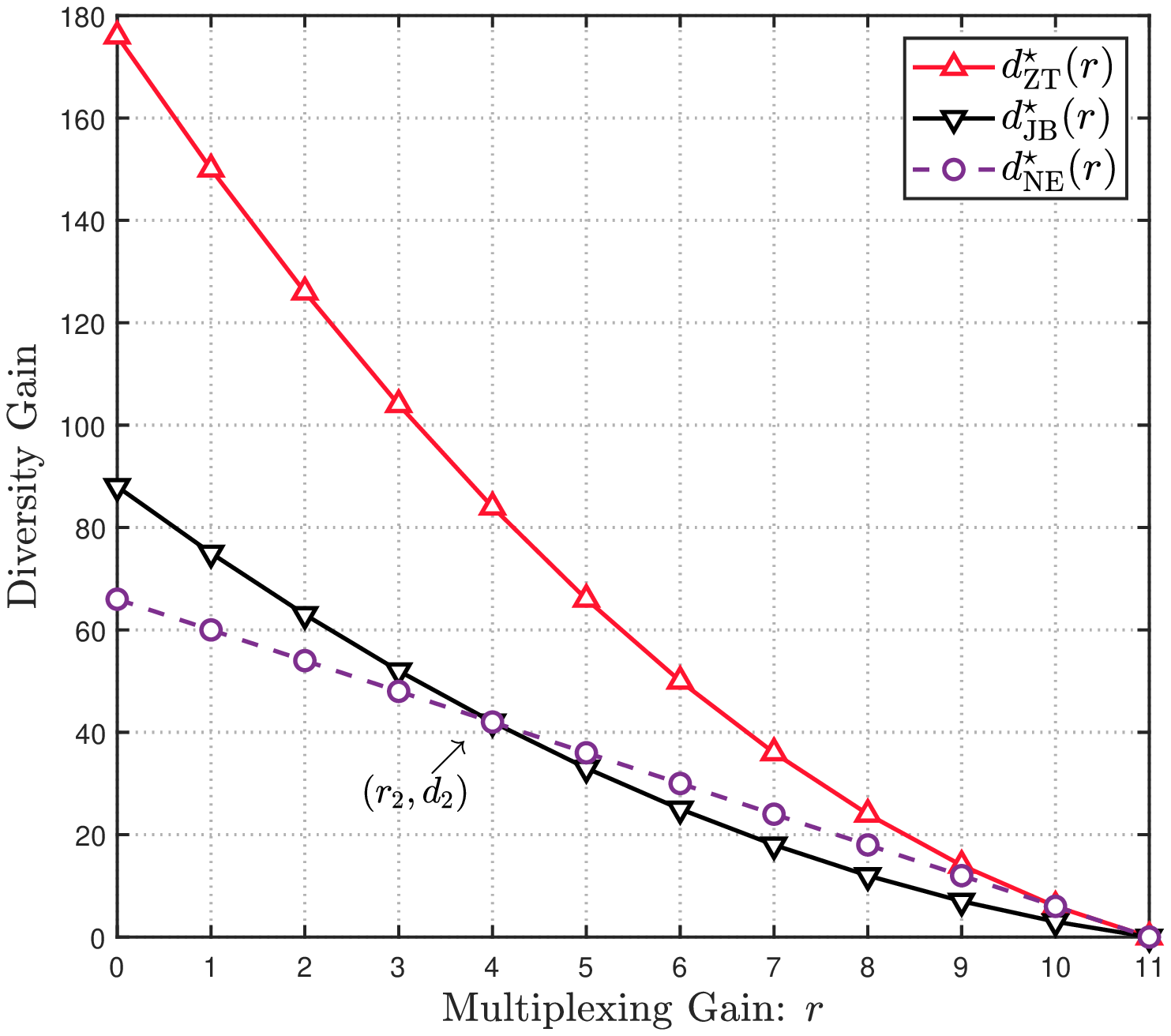}\\
  \caption{Comparisons of the optimal DMT curve for $16\times11\times\infty$ MIMO channel.}\label{figure: d-compare-16-11}
  \end{center}
\end{figure}

To numerically compare the optimal DMT curves for different channels in Table~\ref{tab1}, Figures~\ref{figure: d-compare-16-4} and \ref{figure: d-compare-16-11} depict these DMT curves of $16\times4$ and $16\times11$ MIMO channels when $l \geq l_{\textnormal{th}}$. For any multiplexing gain $r$, we can see that $d^\star_\textnormal{ZT}(r)$ is always large than $d^\star_\textnormal{JB}(r)$ and $d_\textnormal{NE}^\star(r)$. This implies that the optical signals usually suffer more severe fading than their traditional RF counterparts.

In Figures~\ref{figure: d-compare-16-4} and \ref{figure: d-compare-16-11} we also observe that, even though $d_\textnormal{NE}^\star(r)$ and $d_\textnormal{JB}^\star(r)$ both characterize optimal diversity gain, these two curves are fundamentally different. $d_\textnormal{JB}^\star(r)$ is derived by only considering the fact that input signal is real, and then simply putting a half factor into the traditional RF capacity expression. Here, our derived $d_\textnormal{NE}^\star(r)$ is obtained by taking the unique inputs constraints into consideration. 
Comparing these two figures, $d_\textnormal{NE}^\star(r)$ tends to match $d_\textnormal{ZT}^\star(r)$ as $\nr$ decreases, which is caused by the fact that $d_\textnormal{NE}^\star(r)$ approaches $(\nt-r)(\nr-r)$ as $\nr$ decreases. For specific $\nr$, $d_\textnormal{NE}^\star(r)$ may intersect with $d_\textnormal{JB}^\star(r)$ at some point $(r_2,d_2)$, and when $r < r_2$, $d_\textnormal{NE}^\star(r)$ is strictly less than $d_\textnormal{JB}^\star(r)$, while $d_\textnormal{NE}^\star(r)$ is strictly larger than $d_\textnormal{JB}^\star(r)$ when $r > r_2$.

\begin{figure}
  \begin{center}
  \includegraphics[width=3.5in]{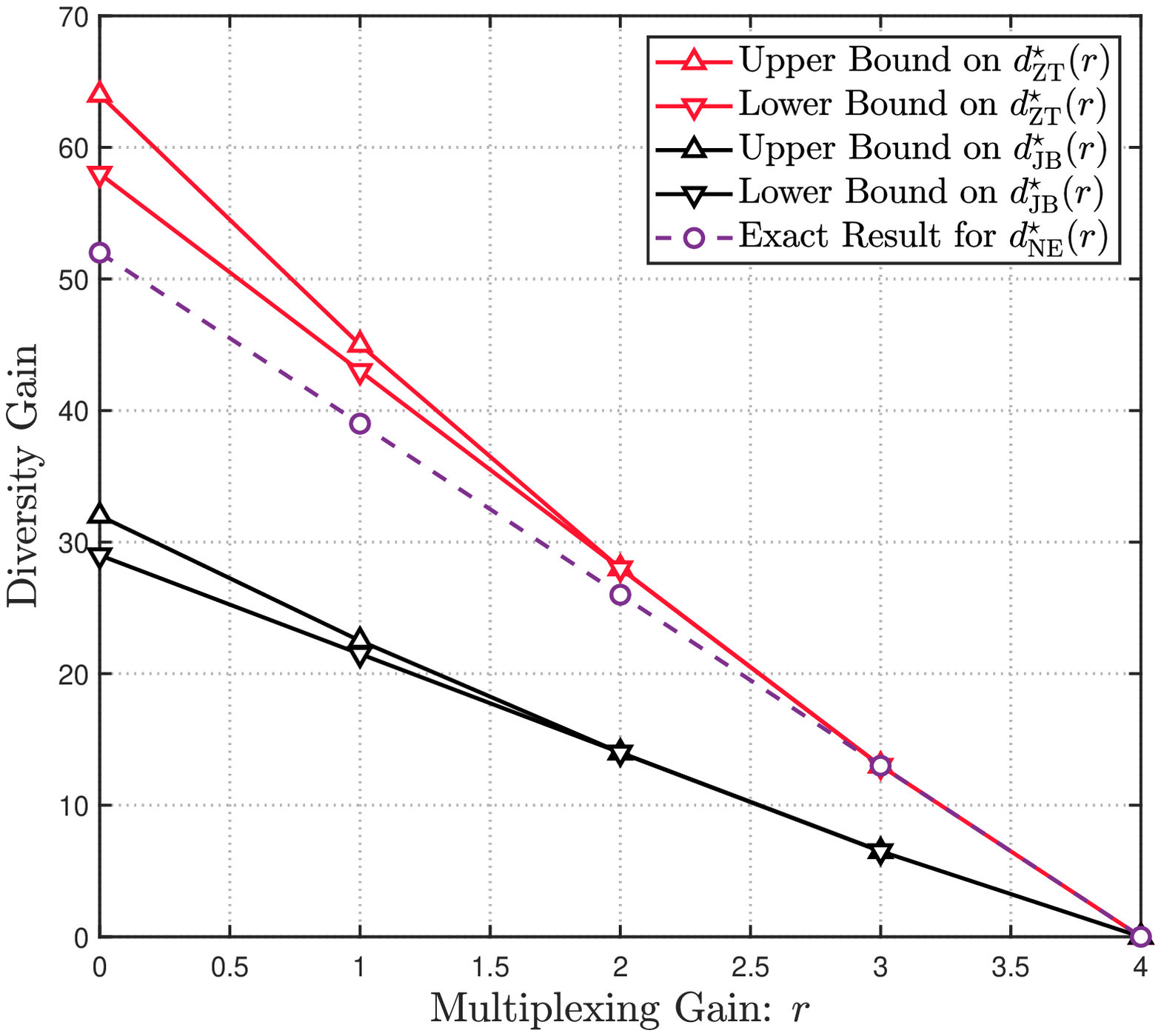}\\
  \caption{Comparisons of the optimal DMT curve for $16\times4\times15$ MIMO channel.}\label{figure: lth 15}
  \end{center}
\end{figure}

\begin{figure}
  \begin{center}
  \includegraphics[width=3.5in]{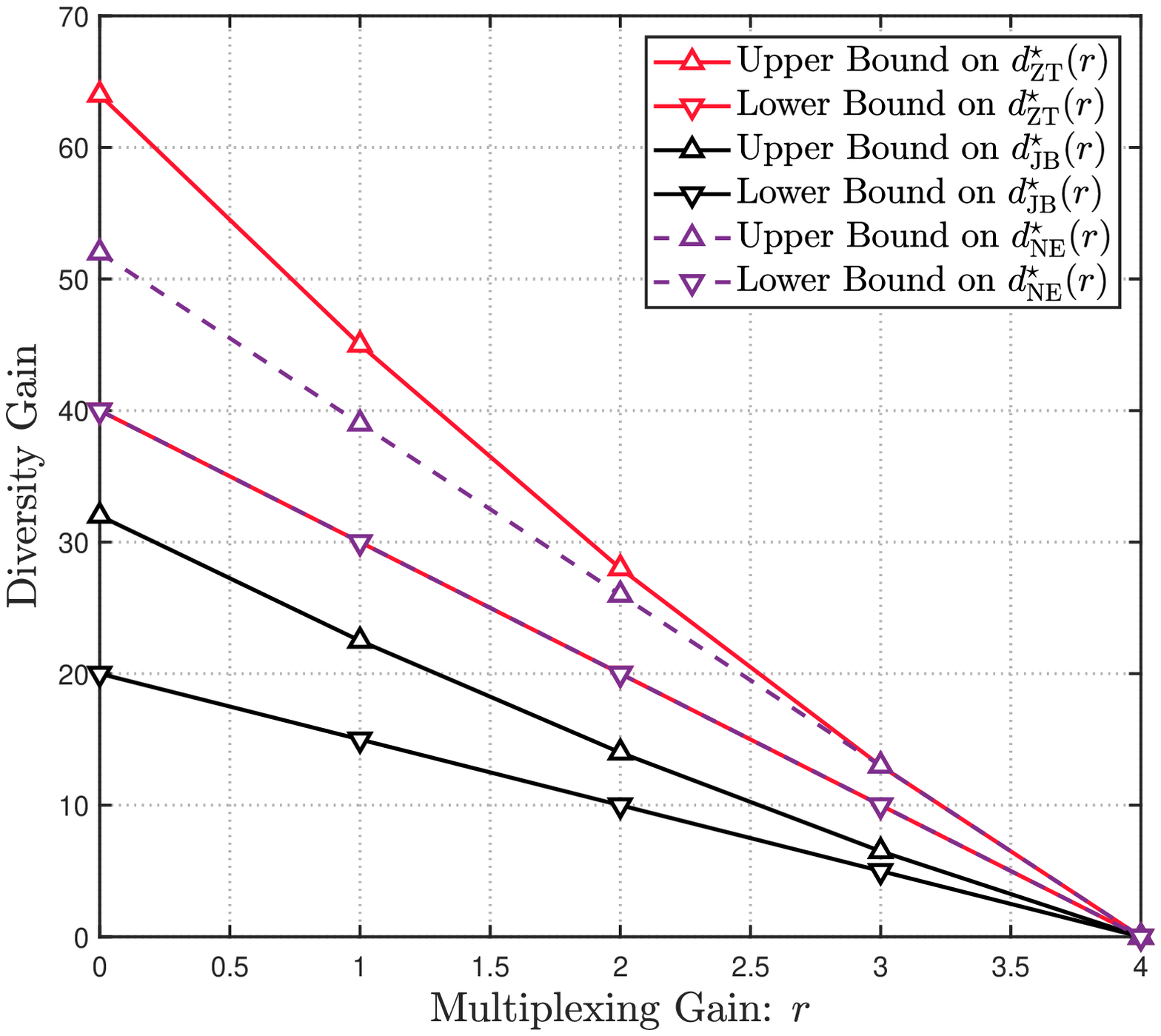}\\
  \caption{Comparisons of the optimal DMT curve for $16\times4\times10$ MIMO channel.}\label{figure: lth 10}
  \end{center}
\end{figure}

Moreover, we numerically study the variation of $d^\star(r)$ when $l$ decreases for a $16\times4$ MIMO channel. Recall that $l^{\textnormal{ZJ}}_{\textnormal{th}}$ denotes the block length threshold for the previous literature \cite{Lizhong_Zheng2003,Jaiswal2019}. Hence, we have $l^{\textnormal{NE}}_{\textnormal{th}}=13$ and  $l^{\textnormal{ZJ}}_{\textnormal{th}}=19$. We select three different block lengths: $(1)\ l=\infty$, $(2)\ l=15$, $(3)\ l=10$, as shown in Figures~\ref{figure: d-compare-16-4},~\ref{figure: lth 15} and~\ref{figure: lth 10}. Note that as $l$ decreases, the gap between the upper and lower bounds on $d^{\star}(r)$ increases. However, our derived $d_\textnormal{NE}^\star(r)$ has the slowest uncertainty growth rate compared with the existing results. Since when $l$ decreases from $\infty$ to $15$, the bounds $d_\textnormal{ZT}^\star(r)$ and $d_\textnormal{JB}^\star(r)$ become very loose. In contrast, our derived $d_\textnormal{NE}^\star(r)$ is still tight until $l=10$.

\section{Conclusion Remarks}
\label{sec:conclusion}
This paper investigates three different block fading optical wireless communication channels and derives the optimal DMT under both optical peak- and average-power input constraints. We first establish an upper bound on the optimal diversity gain by characterizing the outage diversity gain. Then by analyzing the error probability of a random coding scheme, we establish a new lower bound. It turns out that these two bounds are close, hence give good approximations on the optimal diversity gain. In fact, at a large block length regime, these bounds match, thus we can precisely characterize the optimal diversity gain. Our derived DMT results are fundamentally different from their counterparts in traditional RF channels. These differences are due to the unique input constraints in the optical wireless channels. From the perspective of a practical $\nt \times \nr$ OWC system design, our results imply at a specific multiplexing order $r$, letting the numbers of transmit and receive antennas satisfy $\lfloor {\nt+1+r} \rfloor = 2\nr$ may achieve the best performance in terms of diversity gain.

\appendices


\section{Derivation of Instantaneous Capacity Bounds}
\label{app: proof of mutual information bounds}

\subsection{Proof of Proposition~\ref{proposition: mutual information bounds}}
Recall $\vect{X}_j = \trans{[\mat{X}_{j1},\mat{X}_{j2},\ldots,\mat{X}_{ j\nt}]}$, it is straightforward to get
\begin{IEEEeqnarray}{rCl}
    \E{{\mat{X}}_{ji}}
    &=&\amp\left(\frac{1}{\mu}-\frac{e^{-\mu}}{1-e^{-\mu}}\right),  \quad \forall i \in \{1,2,\ldots, \nt \}  \label{expection of Xi}
\end{IEEEeqnarray}
and 
\begin{IEEEeqnarray}{rCl}
    \hh(\mat{X}_{ji}) =\log{\amp}+ 1 -\frac{\mu e^{-\mu}}{1-e^{-\mu}} - \log \frac{\mu}{1-e^{-\mu}}. \label{eq:hx}
\end{IEEEeqnarray}
Next, we define a new random vector $\widetilde{\vect{X}}_j = \trans{[\widetilde{\mat{X}}_{j1}, \widetilde{\mat{X}}_{j2},\ldots, \widetilde{\mat{X}}_{j\nt}]}$, whose entries are i.i.d. according to Gaussian distribution with expectation $0$, and variance
\begin{equation}
\Var{\widetilde{\mat{X}}_{ji}} = \frac{1}{2\pi e}2^{2\hh({\mat{X}}_{ji})}. \label{eq:varxtilde}
\end{equation}
By the Gaussian differential entropy formula, we can immediately show
\begin{IEEEeqnarray}{c}
   \hh(\widetilde{\mat{X}}_{ji}) =  \hh(\mat{X}_{ji}), \ \forall i \in \{1,\ldots,\nt\}.  \label{eq:hxtilde}
\end{IEEEeqnarray}
For a given channel realization $\mat{H} = \matt{H}$, apply the generalized EPI in \cite{Ram_Zamir1993}, and we have
\begin{IEEEeqnarray}{rCl}
    \hh(\matt{H}\vect{X}_j)
    &\geq& \hh(\matt{H}\widetilde{\vect{X}}_j) \\
    &=& \frac{\nr}{2}\log{\left(2\pi e\left|\matt{H}\matt{K}_{\widetilde{\vect{X}}_j\widetilde{\vect{X}}_j} \trans{\matt{H}}\right|^\frac{1}{\nr}\right)}, \label{generalized EPI}
\end{IEEEeqnarray}
where $\matt{K}_{\widetilde{\vect{X}}_j\widetilde{\vect{X}}_j}$ denotes the covariance matrix of $\widetilde{\vect{X}}_j$ which is diagonal with its $i$th entry 
\begin{equation}
\left[\matt{K}_{\widetilde{\vect{X}}_j\widetilde{\vect{X}}_j}\right]_{ii} = \Var{\widetilde{\mat{X}}_{ji}},\quad \forall i \in \{1,2,\ldots,\nt\}.  \label{eq:varxtilde2}
\end{equation}
Substituting~\eqref{eq:hx} into~\eqref{eq:varxtilde},~\eqref{eq:varxtilde2} and~\eqref{generalized EPI}, we have
\begin{equation}
    \hh\left(\matt{H}{\vect{X}}_j\right) \geq \frac{\nr}{2}\log{\left(2\pi e \left|\frac{\sigma_n^2}{2\pi \alpha^2 e} {\mathrm{T}({\alpha},{\nt})}^2  (\OSNR)^2 \matt{H}\trans{\matt{H}}\right|^\frac{1}{\nr}\right)}, \label{new R}
\end{equation}
where
\begin{equation}
    \mathrm{T}({\alpha},{\nt}) = 2 \frac{1-e^{-\mu}}{\mu} 2^{-\frac{\mu e^{-\mu}}{1-e^{-\mu}}},  \label{definition of T}
\end{equation} 
with $\mu$ satisfying~\eqref{eq:mu2}.

Now we lower-bound the instantaneous capacity as
\begin{IEEEeqnarray}{rCl}
    \II\left(\vect{X}_j;\vect{Y}_j\middle|\mat{H}\right)
    &=&\hh\left(\mat{H}\vect{X}_j+\vect{Z}_j\middle|\mat{H}\right)-\hh\left(\vect{Z}_j\right) \label{independence between H and z}\\
    &\geq&  \frac{1}{2}\log{\left(2^{2\hh\left(\mat{H}\vect{X}_j\right)}+2^{2\hh\left(\vect{Z}_j\right)}\right)}-\hh\left(\vect{Z}_j\right)\label{EPI}\\
    &=& \frac{1}{2}\log{\left(1+\frac{2^{2\hh\left(\mat{H}{\vect{X}}_j\right)}}{\left(2\pi e\sigma_n\right)^{\nr}}\right)}\label{S} \\
    &\geq&  \frac{1}{2} \log \left(1+L_{l} (\OSNR)^{2\nr}\left|\mat{H}\trans{\mat{H}}\right|\right)\label{EPI2},
\end{IEEEeqnarray}
where \eqref{independence between H and z} follows by the independence between $\mat{H}$ and $\vect{Z}_j$, \eqref{EPI} by applying the EPI \cite{2014EPI}, and \eqref{EPI2} by substituting~\eqref{new R} into~\eqref{S}. The proof is concluded.

\subsection{Proof of Proposition~\ref{proposition: mutual information bounds2}}
At high OSNR, by the asymptotic upper bound in \cite[Theorem~21]{longguang_li2020}, we have
\begin{IEEEeqnarray}{C}
    \II\left(\vect{X}_j;\vect{Y}_j\middle|\mat{H}\right) \leq
    \nr \log \amp +\frac{1}{2} \log \left( \frac{\mathsf{V}_{\mat{H}}^2}{(2\pi e\sigma_{n})^{\nr}} \right), \label{capacity original upper bound}
\end{IEEEeqnarray}
where $\mathsf{V}_\mat{H}=\sum_\set{U} |\mathrm{det}\mat{H}_\set{U}|$ with $\matt{H}_{\set{U}}$ denoting the $\nr \times \nr$ submatrix of $\matt{H}$ indexed by set
$\set{U}$. 

We further upper bound the RHS of \eqref{capacity original upper bound} as
\begin{IEEEeqnarray}{rCl}
        \nr \log  \amp +\frac{1}{2} \log \left( \frac{\mathsf{V}_{\mat{H}}^2}{(2\pi e \sigma_{n})^{\nr}} \right) 
    &\leq& \nr \log \amp +\frac{1}{2} \log \left( \frac{{{\nt}\choose{\nr}} \left(\sum_\set{U} |\mat{H}_\set{U}|^2\right)}{(2\pi e \sigma_{n})^{\nr}} \right) \label{square mean}\\
    &=& \nr \log \amp +\frac{1}{2} \log \left( \frac{{{\nt}\choose{\nr}} |\mat{H}\trans{\mat{H}}|}{(2\pi e\sigma_{n})^{\nr}} \right), \label{CauchyBinet formula}
\end{IEEEeqnarray}
where \eqref{square mean} holds by Cauchy–Schwarz inequality, and \eqref{CauchyBinet formula} by Cauchy–Binet formula: $\sum_\set{U} |\mat{H}_\set{U}|^2 = |\mat{H}\trans{\mat{H}}|$. The proof is concluded.

\section{Derivation of Outage Probability Bounds}
\subsection{Preliminary}
\label{proof of f(a)}
\subsubsection{Decomposition on Channel Matrix}
Given the channel matrix $\matt{H}\in\set{R}_{+}^{\nr\times\nt}$, we first decompose it by LQ factorization  
\begin{IEEEeqnarray}{rCl}
    \matt{H}=\matt{LQ}, \label{LQ factorization}
\end{IEEEeqnarray}
where $\matt{L}\in\set{R}^{\nr\times\nr}$ is lower-triangular and $\matt{Q}\in\set{R}^{\nr\times\nt}$ is orthogonal satisfying $\trans{\matt{QQ}} = \mat{I}_{\nr\times\nr}$. Combined with \eqref{LQ factorization}, we can rewrite $\trans{\matt{HH}}$ as
\begin{IEEEeqnarray}{rCl}
    \trans{\matt{HH}}=\trans{\matt{LL}}. \label{hh e1}
\end{IEEEeqnarray}
Then perform eigenvalue decomposition on $\trans{\matt{HH}}$, we obtain
\begin{IEEEeqnarray}{rCl}
    \trans{\matt{HH}}=\matt{U}\Lambda\trans{\matt{U}},\label{hh e2}
\end{IEEEeqnarray}
where $\matt{U}\in\set{R}^{\nr\times\nr}$ is orthogonal satisfying $\trans{\matt{UU}} = \mat{I}_{\nr\times\nr}$ and $\Lambda\in\set{R}^{\nr\times\nr}$ is diagonal with $[\Lambda]_{ii}=\lambda_i, \ \forall i=\{1,\cdots,\nr\}$.
Combining \eqref{hh e1} with \eqref{hh e2}, we can decompose $\trans{\matt{LL}}$ as 
\begin{IEEEeqnarray}{rCl}
    \matt{L}\trans{\matt{L}}=\left(\matt{U}\Lambda^{\frac{1}{2}}\right) \trans{\left(\matt{U}\Lambda^{\frac{1}{2}}\right)}. \label{LL decompose}
\end{IEEEeqnarray}
Then we have $\matt{L}=\matt{U}\Lambda^{\frac{1}{2}}$. Further substitute $\matt{L}$ into \eqref{LQ factorization}, and we have 
\begin{IEEEeqnarray}{rCl}
    \matt{H} = \matt{U} \Lambda^{\frac{1}{2}} \matt{Q}.
\end{IEEEeqnarray}
Denote $\matt{D} \eqdef \Lambda^{\frac{1}{2}}$, and rewrite $\lambda_i$ in terms of $a_i,\ \forall i =\{1,\cdots,\nr\}$, then we get
\begin{equation}
\matt{D}=\operatorname{diag}[(\OSNR)^{-\frac{a_{1}}{2}}, \ldots, (\OSNR)^{-\frac{a_{\nr}}{2}}].
\end{equation}

\subsubsection{Derivation of Ep.~\eqref{f(a) definition} }
By \cite{Alan1988}, we have
\begin{IEEEeqnarray}{rCl}
    f(a)  
    & = & \xi\ [\log(\OSNR)]^{\nr} \times \left(\prod_{i=1}^{\mathrm{n}_{\mathrm{R}}} (\OSNR)^{-\frac{\left(\nt-\nr+1\right)}{2} a_{i}} \right) 
 \nonumber \\
    &&  \times \prod_{i<j}\left|(\OSNR)^{-a_{i}}-(\OSNR)^{-a_{j}}\right| \times \int_{\matt{V}_{\nr, \nr}} \int_{\matt{V}_{\nr, \nt}} {f}(\matt{UDQ}) d \matt{Q} d \matt{U},
    \label{pdf of a}
\end{IEEEeqnarray}
where $\xi$ is a normalization constant. 

Form the definition in \eqref{d_out definition}, we observe that the outage diversity order $d_\textnormal{out}(r)$ is fully determined by the corresponding $\OSNR$ exponent of outage probability. Note that
\begin{IEEEeqnarray}{c}
    \lim_{\OSNR \rightarrow \infty} \frac{\log \left(\xi[\log( \OSNR)]^{\nr}\right)}{\log (\OSNR)}=0. \label{eq:osnrexp1}
\end{IEEEeqnarray}
Thus, at high OSNR, we have 
\begin{equation}
\xi [\log (\OSNR)]^{\nr} \doteq \textnormal{OSNR}^0.
\end{equation} 

Second, note that the term $\left|(\OSNR)^{-a_i}-(\OSNR)^{-a_j}\right|$ is only determined by the smaller exponent at high $\OSNR$, and thus we can simplify the following product term as
\begin{IEEEeqnarray}{rCl}
    \prod_{i<j}\left|(\OSNR)^{-a_{i}}-(\OSNR)^{-a_{j}}\right| 
    & \doteq & \prod_{i=1}^{\nr} (\OSNR)^{-(i-1)a_i}. \label{eq:osnrexp2}
\end{IEEEeqnarray}

Substituting~\eqref{eq:osnrexp1} and~\eqref{eq:osnrexp2} into~\eqref{pdf of a}, the proof is concluded. 


\subsection{Proof of Theorem~\ref{proposition-outage probability bounds}}
\label{app-proof of new bounds on outage probability}
\subsubsection{Upper Bound}
\label{sec:upperbound}
To prove the new bounds on outage probability in Theorem~\ref{proposition-outage probability bounds}, we employ the bounds obtained in Proposition~\ref{proposition: mutual information bounds}. We first prove the upper bound on outage probability, and then prove the lower bound. 


Denote $R=r\log \OSNR$. Substitute~\eqref{lower bound on mutual information} into the RHS of~\eqref{Pout definition}, and we obtain 
\begin{IEEEeqnarray}{rCl}
        \mathrm{P}_\textnormal{out}(\OSNR) 
    &\leq& \mathrm{P} \Biggl[\frac{1}{2} \log \left(1+L_l (\OSNR)^{2\nr}\left|\mat{H}\trans{\mat{H}}\right|\right) \leq r\log(\OSNR)\Biggr] \\
    &=& \mathrm{P} \Biggl[  1+L_l (\OSNR)^{2\nr}\left|\mat{H}\trans{\mat{H}}\right| \leq {\OSNR}^{2r}\Biggr].
    \label{outgae probability upper bound e2}
\end{IEEEeqnarray}

Since ${\lambda}_i = (\OSNR)^{-{a}_i}$, $\forall i\in \{1,\ldots,\nr\}$, are the the eigenvalues of $\mat{H}\trans{\mat{H}}$, we have
\begin{equation}
\left|\mat{H}\trans{\mat{H}}\right| = \prod_{i=1}^{\nr}{\lambda}_i = {\OSNR}^{-\sum_{i=1}^{\nr}{a}_i}. \label{eq:det}
\end{equation}

Substituting~\eqref{eq:det} into~\eqref{outgae probability upper bound e2}, we get
\begin{IEEEeqnarray}{rCl}
        \mathrm{P}_{\textnormal{out}}(\OSNR) 
    &\leq& \mathrm{P}\Biggl[1+L_l (\OSNR)^{2 \nr - \sum_{i=1}^{\nr} {a}_i}  \leq (\OSNR)^{2 r}\Biggr] \\
    &\doteq& \mathrm{P}\Biggr[ (\OSNR)^{\left(2\nr-\sum_{i=1}^{\nr}{a}_i\right)^{+}} \leq (\OSNR)^{2 r}\Biggr] \label{eq:+++}\\
    &=&\mathrm{P}\Biggl[(2\nr-\sum_{i=1}^{\nr} {a}_i)^{+} \leq 2r\Biggr]\\
    &=& \int_{\set{A}} f(a) d a, \label{outage P upper bound relation with set A}
\end{IEEEeqnarray}
where~\eqref{eq:+++} follows by the fact that $1+ c\cdot {\OSNR}^b \doteq \OSNR^{b^+} $, and \eqref{outage P upper bound relation with set A} by the definition in~\eqref{set A definition}. The proof is concluded.


\subsubsection{Lower Bound}
\label{app-proof of new bounds on outage probability2}

Substituting \eqref{upper bound on mutual information} into the RHS of~\eqref{Pout definition}, and following the similar arguments as in Section~\ref{sec:upperbound}, we have
\begin{IEEEeqnarray}{rCl}
        \mathrm{P}_\textnormal{out}(\OSNR) 
    &\geq& \mathrm{P}\Biggl[ \frac{1}{2} \log \left(L_u(\OSNR)^{2\nr}\left|\mat{H}\trans{\mat{H}}\right|\right) \leq r\log(\OSNR) \Biggr]\\
    &=& \mathrm{P}\Biggr[L_u (\OSNR)^{2 \nr - \sum_{i=1}^{\nr}{a}_i}  \leq  (\OSNR)^{2 r}\Biggr] \\
    &\doteq& \mathrm{P}\Biggr[ (\OSNR)^{\left(2\nr-\sum_{i=1}^{\nr}{a}_i\right)} \leq (\OSNR)^{2 r}\Biggr] \label{eq:+++2}\\
    &\doteq& \mathrm{P}\left[ 2 \nr -\sum_{i=1}^{\nr} {a}_i \leq 2r\right]\label{probability of set B} \\
    &=& \int_{\set{B}} f(a) d a \label{outage P upper bound relation with set B}.
\end{IEEEeqnarray}
The proof is concluded.

\section{Proof of Eq.~\eqref{condition error probability2}}
\label{sec-Proof of Proposition random coding analysis}
Denote the codebook as $\{\matt{X}(0),\matt{X}(1),\ldots,\matt{X}({\OSNR}^{lr}-1)\}$. Suppose the sent codeword is $\matt{X}(0)$, and we consider the event when the ML decoder decides erroneously in favor of $\matt{X(1)}$. This event occurs only if the projection distance of $\mat{Y(0)}-\matt{H}\matt{X(0)}$ on the direction of $\matt{H}\matt{X(1)}-\matt{H}\matt{X(0)}$ is larger than $d_{0 1}/2$, where $d_{0 1}=\parallel\matt{H}\matt{X(1)}-\matt{H}\matt{X(0)} \parallel_\mathsf{F}$ is the distance between $\matt{H}\matt{X(1)}$ and $\matt{H}\matt{X(0)}$. The probability on the occurrence of this event is\footnote{Without loss of generality, for any $\matt{X}\in\set{R}^{m\times n}$ and $\matt{Y}\in\set{R}^{m\times n}$, $\left<\matt{X},\matt{Y}\right>=\sum_{i=1}^{m}\sum_{j=1}^{n}x_{i j}y_{i j }$.}  
\begin{align}
    &\mathrm{P}\left(\matt{X(0)} \to \matt{X(1)} | \mat{H}=\matt{H}\right) \notag\\
    &\qquad=\mathrm{P}\Biggl(\frac{\left<\bm{\mat{Y(0)}}-\matt{H}\matt{X(0)},  \matt{H}\matt{X(1)}-\matt{H}\matt{X(0)}\right>} {\parallel\matt{H}\matt{X(1)}-\matt{H}\matt{X(0)} \parallel_\mathsf{F}} 
   \geq \frac{1}{2} \parallel\matt{H}\matt{X(1)}-\matt{H}\matt{X(0)} \parallel_\mathsf{F}\Biggr).
\end{align}
Since $\matt{X(0)}$ is sent, and $\bm{\mat{Z(0)}} = \bm{\mat{Y(0)}} -\matt{H}\matt{X(0)}$, then the projection of $\bm{\mat{Y(0)}}-\matt{H}\matt{X(0)}$ on $\matt{H}\matt{X(1)}-\matt{H}\matt{X(0)}$ is still a Gaussian variable. In the following, we use the term $\frac{\rv{z}} {\parallel\matt{H}\matt{X(1)}-\matt{H}\matt{X(0)} \parallel_\mathsf{F}}$ to denote the projection, where $\rv{z}$ is a Gaussian variable with expectation $0$ and variance $\parallel\matt{H}\matt{X(1)}-\matt{H}\matt{X(0)} \parallel_\mathsf{F}^2 {\sigma_n}^2$. Thus we further upper-bound the error probability as
\begin{IEEEeqnarray}{rCl}
\mathrm{P}\left(\matt{X(0)} \to \matt{X(1)} | \mat{H}=\matt{H}\right)
    &=& \mathrm{P}\left(\frac{\rv{z}} {\parallel\matt{H}\matt{X(1)}-\matt{H}\matt{X(0)} \parallel_\mathsf{F}}\geq \frac{1}{2} \parallel\matt{H}\matt{X(1)}-\matt{H}\matt{X(0)} \parallel_\mathsf{F}\right) \nonumber \\
    \\
    &=& Q\left(\sqrt{ \frac{\parallel\matt{H}\matt{X(1)}-\matt{H}\matt{X(0)} \parallel_\mathsf{F}^2}{4 {\sigma_n}^2} }\right)\\
    &\leq& \exp\left(-\frac{\parallel\matt{H}\matt{X(1)}-\matt{H}\matt{X(0)} \parallel_\mathsf{F}^2}{8 {\sigma_n}^2}\right), \label{conditional pair error probability}
\end{IEEEeqnarray}
where~\eqref{conditional pair error probability} holds because $Q(t) \leq 1/2\exp(-t^2/2)$.    

Note that 
\begin{align}
    \parallel\matt{H}\matt{X(1)}&-\matt{H}\matt{X(0)} \parallel_\mathsf{F}^2 =\sum_{k=1}^{\nr}{\sum_{j=1}^{l}{|\sum_{i=1}^{\nt}[\matt{X}_{ij}(1)-\matt{X}_{ij}(0)]\rvv{h}_{ki}|^2}}, \label{distance}
\end{align}
where $\matt{X}_{ij}(\cdot) = [\matt{X}(\cdot)]_{ij}$, and $\rvv{h}_{ki} = [\matt{H}]_{ki}$.
Setting ${\Omega}_j = \trans{[\matt{X}_{1j}(1)-\matt{X}_{1j}(0),\cdots,\matt{X}_{\nt j}(1)-\matt{X}_{\nt j}(0)]}$, we rewrite \eqref{distance} as
\begin{align}
    \parallel\matt{H}\matt{X(1)}-\matt{H}\matt{X(0)} \parallel_\mathsf{F}^2 =\sum_{j=1}^{l} {\trans{\Omega_j}}\trans{\matt{H}}\matt{H}{\Omega}_j \label{distance 2}.
\end{align}
By SVD, we decompose $\trans{\matt{H}}\matt{H}=\matt{V}\Lambda \trans{\matt{V}}$, where $\Lambda =$ $\mathrm{diag}[\lambda_1,...,\lambda_{\nt}]$. Define $\beta_j=[{\beta}_{j 1},\cdots,{\beta}_{j \nt}]$, and let $\beta_j = \trans{{\Omega}_{j}} \matt{V}$. We can rewrite~\eqref{distance 2} as
\begin{align}
    \parallel\matt{H}\matt{X(1)}-\matt{H}\matt{X(0)} \parallel_{\rm F}^2 =\sum_{j=1}^{l} \sum_{i=1}^{\nt} \lambda_i \beta_{j i}^2. \label{projection distance}
\end{align}
Substituting \eqref{projection distance} into \eqref{conditional pair error probability}, we have
\begin{IEEEeqnarray}{rCl}
    \mathrm{P}\left(\matt{X(0)} \to \matt{X(1)} | \mat{H}=\matt{H}\right)
    \leq \exp\left(-\frac{1}{8{\sigma_n}^2} \sum_{j=1}^l \sum_{i=1}^{\nt} \lambda_i {\beta}_{j i}^2\right).
\end{IEEEeqnarray}
Averaging over the ensemble of truncated exponential random codes, we get 
\begin{IEEEeqnarray}{rCl}
\mathrm{P}(\bm{\mat{X(0)}}\to\bm{\mat{X(1)}}|\mat{H}=\matt{H})
    &\leq& \E[\bm{\beta}_{j i}]{\exp\biggl(-\frac{1}{8{\sigma_n}^2} \sum_{j=1}^l \sum_{i=1}^{\nt} \lambda_i \bm{\beta}_{j i}^2\biggr)}\\
    &=& \prod_{j=1}^{l} \E[\bm{\beta}_{j}]{\exp(-\frac{1}{8{\sigma_n}^2} \sum_{i=1}^{\nt} \lambda_{i} \bm{\beta}_{j i}^2)} \label{expection 1}\\
    &=&\mathrm{det}\left(\mat{I}+\frac{g(\alpha,\nt)}{2}(\OSNR)^2 \matt{H}\trans{\matt{H}}\right)^{-\frac{l}{2}}, \label{e6}
\end{IEEEeqnarray}
where \eqref{expection 1} follows by the independence between vectors $\bm{\beta}_{p}$ and $\bm{\beta}_{q}, \,\,\forall p \ne q$, and~\eqref{e6} by the derivation in the following Appendix~\ref{app:Proposition-pair error probility proof}.

Note that at transmit rate $\R=r\log(\OSNR)$, we have in total $(\OSNR)^{l r}$ codewords. Then applying the union bound, the decoded error probability can be upper-bounded as
\begin{IEEEeqnarray}{rCl}
\mathrm{P}(\mathrm{error}|\mat{H}=\matt{H})
    &\leq& (\OSNR)^{lr} \mathrm{det}\left(\mat{I}+\frac{g(\alpha,\nt)}{2}(\OSNR)^2 \matt{H}\trans{\matt{H}}\right)^{-\frac{l}{2}}\\
    &=& (\OSNR)^{lr} \prod_{i=1}^{\nr}\left(1+\frac{g(\alpha,\nt)}{2}(\OSNR)^2\lambda_i\right)^{-\frac{l}{2}}.\label{condition error probability}
\end{IEEEeqnarray}

\subsubsection{Proof of Eq.~\eqref{e6}}
\label{app:Proposition-pair error probility proof}
Denote the matrix $\matt{V}=[\mathrm{V}_1,\cdots,\mathrm{V}_{\nt}]$, where $\trans{\mathrm{V}_p = [\rvv{v}_{1p},\cdots,\rvv{v}_{\nt p}]}$ is the $p$th column of $\matt{V}$. Recall that $\matt{V}$ is an orthogonal matrix, i.e.,
\begin{IEEEeqnarray}{rCl}    
    \sum_{i=1}^{\nt}\rvv{v}_{i p}\rvv{v}_{i p} &=& 1, \quad \forall p\in\{1,\cdots,\nt\},\\
    \sum_{i=1}^{\nt}\rvv{v}_{i p}\rvv{v}_{i q} &=& 0, \quad \forall p\ne q .\label{orthogonal}
\end{IEEEeqnarray}
Note that
\begin{IEEEeqnarray}{rCl}
    \bm{\beta}_{j p} 
    &=& \Delta\mat{X}_{1 j}\rvv{v}_{1p}+\cdots+\Delta\mat{X}_{\nt j}\rvv{v}_{\nt p} \label{beta-jp}\\
    \bm{\beta}_{j q} 
    &=& \Delta\mat{X}_{1 j}\rvv{v}_{1q}+\cdots+\Delta\mat{X}_{\nt j}\rvv{v}_{\nt q}, \label{beta-jq}
\end{IEEEeqnarray}
where $\Delta\mat{X}_{i j}=\mat{X}_{i j}\bm{(1)}-\mat{X}_{i j}\bm{(0)}$. 

Recall that $\vect{X}_{i j}$, $\forall i\in\{1,\cdots,\nt\}$ and $j\in\{1,\cdots,l\}$, follows i.i.d. truncated exponential distribution in \eqref{exponential distribution2}, and hence it is directly to verify
\begin{IEEEeqnarray}{rCl}
    \E{\bm{\beta}_{j i }} &=& 0, \label{beta mean}\\
    \Var{\bm{\beta}_{j i }} &=& 2 \amp^2 \left( {\frac{1}{\mu^2}-\frac{e^{-\mu}}{(1-e^{-\mu})^2}} \right)\\
    &\triangleq&2 \amp^2 g(\alpha,\nt).\label{beta variance}
\end{IEEEeqnarray}

By the central limit theorem, we can approximate $\bm{\beta}_{j i}$ as a Gaussian random variable with expectation~\eqref{beta mean} and variance~\eqref{beta variance}, and hence by averaging over the distribution of $\bm{\beta}_{j i}$ in~\eqref{expection 1}, we have
\begin{equation}
    \prod_{j=1}^{l} \E[\bm{\beta}_{j}]{\exp(-\frac{1}{8{\sigma_n}^2} \sum_{i=1}^{\nt} \lambda_{i} \bm{\beta}_{j i}^2)}
    =\mathrm{det}\left(\mat{I}+\frac{g(\alpha,\nt)}{2}(\OSNR)^2 \matt{H}\trans{\matt{H}}\right)^{-\frac{l}{2}}. \label{e7}
\end{equation}
The proof is concluded.



\end{document}